\documentclass[reprint,aps,prb,noeprint,notitlepage,nofootinbib,superscriptaddress]{revtex4-2}
\usepackage{amsmath}
\usepackage{amsfonts}
\usepackage{amssymb}
\usepackage{graphicx}
\usepackage{epstopdf}
\usepackage{caption}
\usepackage{subcaption}
\usepackage{comment}
\usepackage{physics}
\usepackage{float}
\usepackage[colorlinks=true]{hyperref}
\usepackage[colorlinks=true]{hyperref}


\input pdfmsym

\pdfmsymsetscalefactor{8}  

\catcode`@=11

\def\@rstraightarrow{\@linehead@type{0 0 m 2 0 l S 2 0 m 0 1.25 l 1 0 l 0 -1.25 l 2 0 l b}{2}}
\def\@lstraightarrow{\@linehead@type{2 0 m 0 0 l S 0 0 m 2 1.25 l 1 0 l 2 -1.25 l 0 0 l b}{2}}

\@arrow@def{rsarrow}\@linecap\@rstraightarrow
\@arrow@def{lsarrow}\@lstraightarrow\@linecap
\@arrow@def{lrsarrow}\@lstraightarrow\@rstraightarrow

\catcode`@=12

\begin{document}
\title{Unidirectional gliding of a cycloidal spin structure by an AC magnetic field}
\author{Dong Hui Han}
\affiliation{Department of Physics, Korea Advanced Institute of Science and Technology, Daejeon 34141, Republic of Korea}
\author{Kyoung-Woong Moon}
\affiliation{Quantum Technology Institute, Korea Research Institute of Standards and Science, Daejeon 34113, Republic of Korea}
\author{Kab-Jin Kim}
\affiliation{Department of Physics, Korea Advanced Institute of Science and Technology, Daejeon 34141, Republic of Korea}
\author{Se Kwon Kim}
\affiliation{Department of Physics, Korea Advanced Institute of Science and Technology, Daejeon 34141, Republic of Korea}


\begin{abstract}
    The dynamics of a cycloidal spin structure driven by an AC magnetic field is theoretically studied in the weak-field limit. A specific model Hamiltonian describing the cycloidal spin structure in a ferromagnetic thin film is constructed, and its dynamics is analyzed using the collective-coordinate approach within the Lagrangian formalism. We demonstrate that the cycloidal spin structure exhibits a unidirectional gliding motion under an AC magnetic field, and an expression for the average velocity is derived as a function of the magnitude, the direction, and the frequency of the AC magnetic field. We compare our theoretical predictions with the results of micromagnetic simulations and identify two resonance frequencies determined by the eigenenergies of the excitation modes. Furthermore, evaluating spin motive forces induced by the dynamics reveals a substantial DC voltage, which may be exploited in energy-harvesting devices utilizing ambient electromagnetic radiation.
\end{abstract}

\maketitle


\section{Introduction}

Topological solitons in the various classes of magnets have received significant attention from both theoretical and experimental perspectives, as certain dynamical properties of the noncollinear spin structures hold great promise for next-generation devices~\cite{KOSEVICH1990117,Braun01022012}. In magnetic systems, the noncollinear spin structures can arise from a wide range of mechanisms such as frustration~\cite{yoshimori,nagamiya1962magnetization}, long-range dipole interactions~\cite{LIU2018495} and Dzyaloshinskii-Moriya interaction (DMI)~\cite{RevModPhys.91.015004}. DMI is an antisymmetric interaction between nearby spins, which was proposed by Dzyaloshinskii to explain a phenomenon called weak ferromagnetism~\cite{dzyaloshinsky1958thermodynamic}. Given spin-orbit coupling, a relativistic interaction associated with orbiting electrons, Moriya developed the argument by extending Anderson's theory~\cite{anderson1959new,moriya1960anisotropic}. The interaction becomes significant in magnetic materials lacking certain symmetries, resulting in a finite DMI upon spatial integration over the system. One representative system is a thin film on a substrate, where the inversion symmetry is broken along the film normal. In the system, various noncollinear spin structures have been predicted and experimentally realized~\cite{RevModPhys.89.025006}. 


One of those structures is a cycloidal spin structure (CSS). The CSS can emerge in a ferromagnetic thin film due to the interfacial DMI even in the presence of an easy-axis anisotropy along the film normal. However, it is required to have the DMI exceeding a critical value, $D_c$, to stabilize the structure on the ferromagnetic system, implying the existence of an incommensurate-commensurate (IC-C) phase transition at $D_c$~\cite{PhysRevLett.31.459,rohart2013skyrmion,kishine2015theory}. If the DMI is insufficient to overcome the penalty from the exchange and the anisotropy energy, the spin configuration favors a uniform alignment along the easy axis. The relation of those interactions determines $D_c$, as well as the modulation of the nonlinear structure above $D_c$. The CSS can be regarded as a chiral magnetic soliton lattice, generally meaning a periodic lattice structure of solitons with a fixed chirality. For example, in the ferromagnetic thin film, the chirality is fixed by the interfacial DMI, and each soliton has a shape of the N\'{e}el-type domain wall. If the system does not have the easy-axis anisotropy, the structure reduces to a chiral helimagnetic state, a noncollinear linear spin structure~\cite{kishine_coherent_2012,PhysRevB.104.224405}. The CSS has been observed in thin film systems and multiferroic materials such as Mn/W~\cite{bode2007chiral,PhysRevLett.101.027201}, Fe/W(110)~\cite{PhysRevB.78.140403,PhysRevLett.103.157201}, TbMnO\textsubscript{3}~\cite{kenzelmann2005magnetic,PhysRevLett.106.047203}, BiFeO\textsubscript{3}~\cite{PhysRevB.84.144404,Park_2014}. 

Another prototypical soliton is a domain wall (DW). The dynamics of the DW in magnetic systems driven by DC magnetic fields has been extensively studied, and firmly established theories successfully account for experimental results~\cite{beach2005dynamics}. Those theories use the collective-coordinate approach introduced by Schryer and Walker on the dynamics of a Bloch DW~\cite{schryer1974motion} and the approach is broadly applied to capture unique features of the dynamics beyond the case of DWs~\cite{clarke2008dynamics}. In the quasi-one-dimensional DW case, the theory uses two collective coordinates, $X(t)$ and $\Phi(t)$, which define the position of the DW and its azimuthal angle, respectively~\cite{slonczewski1972dynamics,kim2023mechanics}. However, the theory can include more variables, and one example is $\lambda(t)$, which represents the width of the DW~\cite{thiaville2002domain}. Although $\lambda(t)$ is often treated as constant due to larger energy cost for excitations than inducing translational motion or rotation~\cite{thiaville2012dynamics}, it plays a crucial role in the presence of an AC magnetic field. According to recent studies, a DW can also move by an AC magnetic field~\cite{moon2017domain,kim2020magnetic} and the essence of one mechanism in the phenomenon is a coupling between the oscillating width of DW and its translational motion~\cite{moon2017domain}.

The structural resemblance of the CSS to an array of N\'{e}el-type DWs motivates us to extend the theoretical frameworks developed for the dynamics of DWs to the CSS with attention to inherent differences. One distinction is that, as DMI increases, the shape of solitons in the CSS deviates from that of DWs and the distance between solitons is reduced, i.e., the area of each domain decreases. Another important difference is that the CSS is a periodic system that includes repulsive interactions between adjacent solitons~\cite{ghosh2017annihilation}, leading the structure to saturate into a new ground state under a DC magnetic field, in contrast to DWs exhibiting uniform motions. However, the CSS can exhibit a unidirectional gliding motion by an AC magnetic field, i.e., the oscillation center of the CSS gradually shifts from its initial position, since DWs with opposite topological charges acquire the same unidirectional average velocity under the AC magnetic field~\cite{moon2017domain}. As we will show through an analysis of the dynamics among excitation modes, the average velocity originates from a mechanism similar to that in DWs and depends on the intensity, direction, and frequency of the AC magnetic field. 

Furthermore, the observation of the unidirectional gliding motion of the CSS under an AC magnetic field inspires an investigation into the spin motive force (SMF). It is well known that the dynamics of noncollinear spin structures can induce effective electromagnetic fields, which is called SMF~\cite{PhysRevLett.93.127204,PhysRevB.82.054410,PhysRevB.77.134407,PhysRevLett.134.056702}. At the interface of a heterostructure where an electric potential emerges due to broken inversion symmetry, the Rashba effect gives rise to SMF, which contains both adiabatic and non-adiabatic contributions~\cite{PhysRevB.87.054403}. In this study, we evaluate the SMF originating from the Rashba effect and find that, under an AC magnetic field, the CSS generates a substantial DC voltage from the non-adiabatic SMF, acting as a magnetic rectifier that converts an AC magnetic field to a DC voltage. This property offers the potential for enhancing the performance of devices harvesting the energy from electromagnetic radiation~\cite{sharma2024nanoscale}.



In this paper, we theoretically study the dynamics of the CSS in a ferromagnetic thin film driven by an AC magnetic field. We begin by identifying the ground state of the model Hamiltonian and investigating excitation modes to describe the system's dynamics. An analytical expression for the average velocity, including only a small number of modes, shows good agreement with micromagnetic simulations. From an analysis of the expression, we demonstrate that the mechanism of the gliding motion is similar to DWs, and the resonance frequencies that maximize the average velocity correspond to the natural frequencies of the two subsystems which behave as two coupled simple harmonic oscillators, respectively. Finally, we evaluate both the adiabatic and non-adiabatic spin motive forces and verify a substantial DC voltage from the Rashba effect, indicating its potential for future spintronic applications.

In \text{Sec.~\ref{sec:model}}, we present a model and its ground state. In \text{Sec.~\ref{sec:sw}}, we find excitation modes for our model. In \text{Sec.~\ref{sec:dynamics}}, we set a proper ansatz including some modes, derive an average velocity equation using the Lagrangian formalism, compare the theoretical results with simulations, and examine the spin motive force related to the gliding motion. We summarize and discuss our results in \text{Sec.~\ref{sec:conclusion}}.

\section{Model and ground states}
\label{sec:model}

The CSS in a quasi-one-dimensional ferromagnetic thin film can be modeled by a Hamiltonian with strong interfacial DMI and easy-axis anisotropy. In the semi-classical approach, due to slow-varying spin dynamics under the low-energy limit, the Hamiltonian density can be expressed in terms of classical fields of magnetization as 
\begin{equation}
    \mathcal{H}=A(\partial_x \mathbf{m})^2-Km_z^2-\mathbf{D}\cdot(\mathbf{m}\times \partial_x\mathbf{m})-M\mathbf{H}\cdot\mathbf{m}\,,
\end{equation}  
 where $\mathbf{m}$ is a unit magnetization vector, 
$A>0$ is an exchange constant, $K>0$ is a first order easy-axis anisotropy constant, $\mathbf{D}=D\hat{\mathbf{y}}$ is a DMI vector with DMI constant $D$, $M$ is a magnetization and $\mathbf{H}$ is an external magnetic field~\cite{Dzyaloshinskii1964rus,rohart2013skyrmion}.

We can find the ground state when $\mathbf{H}=0$ by representing the magnetization unit vector with two angle fields, $\theta(x)$ and $\phi(x)$, as 
\begin{equation}
    (m_x, m_y, m_z)=(\sin{\theta}\sin{\phi},\cos{\theta}, \sin{\theta}\cos{\phi})\,.
\end{equation}
The ground state lies in the $z$-$x$ plane and exhibits a cycloidally modulated spin structure due to the DMI and the easy-axis anisotropy. Here, the ground state is given by~\cite{Yurii,kishine_coherent_2012}
\begin{figure}[t]
    \centering
    \includegraphics[width=0.48\textwidth]{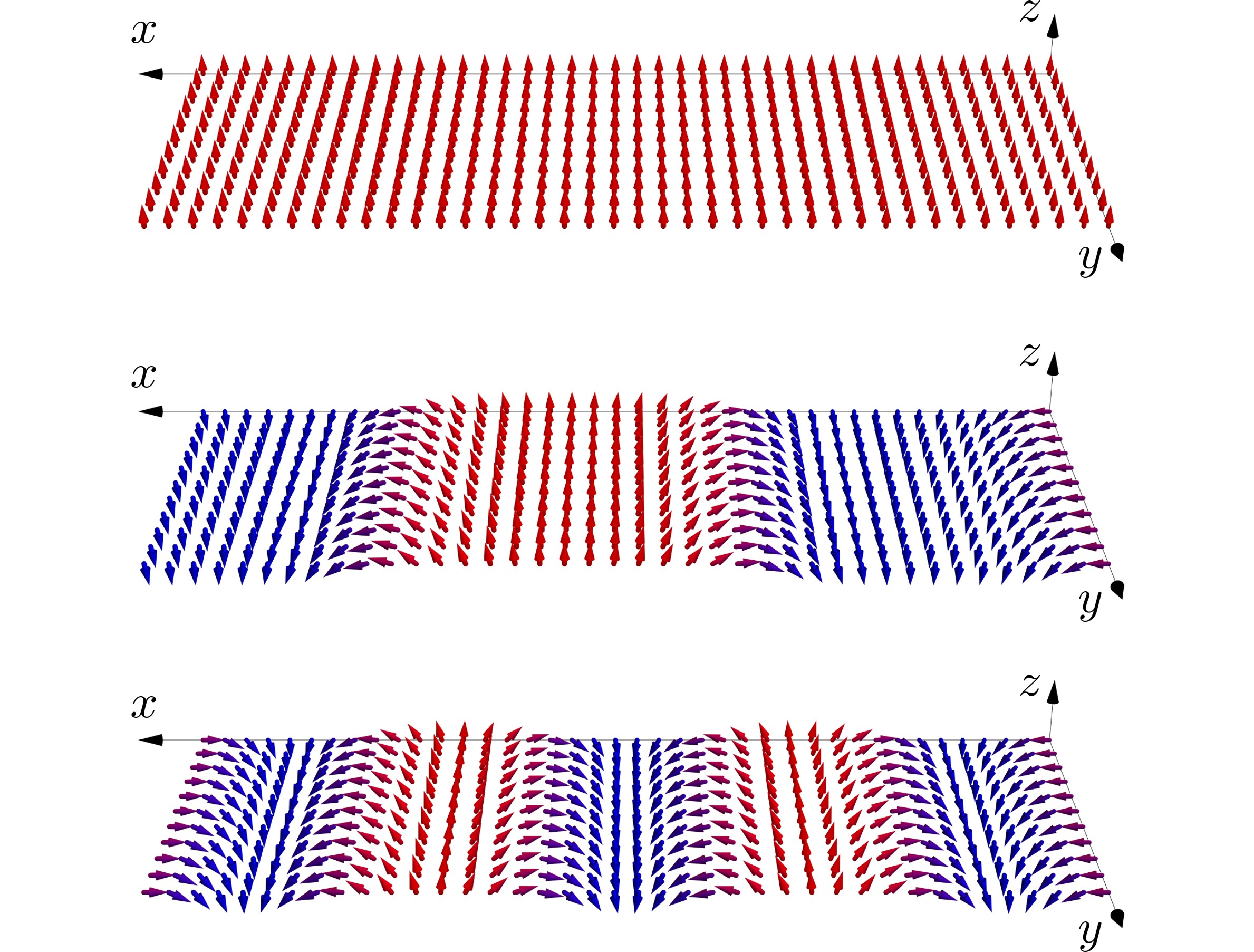} 
    \caption{Simple schematics of the ground states for different DMI constants, $D$. The color of arrows represents $m_z$ and red~(blue) color indicates $m_z=+1\,(-1)$. From top to middle, $D$ increases from $D_1\,(<D_c)$ to $D_2$. From middle to bottom, $D$ increases from $D_2$ to $D_3$, i.e., $D_1<D_c<D_2<D_3$.}
    \label{fig1}
\end{figure}

\begin{equation}
    \theta_0(x)=\frac{\pi}{2}\,, \quad\quad\phi_0(x)=\frac{\pi}{2}+\operatorname{am}\big(\frac{x}{\lambda\kappa},\kappa\big)\,,
\end{equation}
where $\operatorname{am}$ is a Jacobi amplitude function whose period is $L_0=4 \lambda \kappa K(\kappa)$, $\kappa$ is the elliptic modulus, $K(\kappa)$ is the complete elliptic integral of the first kind and $\lambda=\sqrt{A/K}$ [See Appendix~\ref{appa}]. The elliptic modulus, $\kappa$, determines the period of the spin structure, and it is given by
\begin{equation}
    \kappa=\frac{4AE(\kappa)}{\pi \lambda D}=\frac{4E(\kappa)}{\pi D}\sqrt{A K}\,,\label{defkappa}
\end{equation}
where it minimizes the energy of a soliton, $\mathcal{E}_{s}(\kappa)=\int_0^{L_0} dx \mathcal{H}/L_0$, and $E(\kappa)$ is the elliptic integral of the second kind. However, the constraint, $0\leq \kappa \leq1$, for fixed $A$ and $K$ implies the existence of a critical DMI constant, $D_c$, given by 
\begin{equation}
    D_c=\frac{4\sqrt{A K}}{\pi}\,,\label{criticaldmi}
\end{equation}
where $\kappa=1$~\cite{PhysRevLett.103.157201,PhysRevB.78.140403}. In this limit, the period of the CSS diverges, and the IC-C phase transition occurs at $D_c$.
This transition corresponds to a change of the ground state from the uniform ferromagnetic configuration, $\mathbf{m}=\hat{\mathbf{z}}$, to the CSS, a periodic nonlinear spin structure whose period cannot be expressed as a rational multiple of the lattice constant~\cite{PhysRevLett.31.459,chaikin1995principles}. Figure \ref{fig1} illustrates the IC-C phase transition from the ferromagnetic ground state to the CSS, which appears as $D$ exceeds the $D_c$. As $D$ increases beyond $D_c$, the period of the CSS continuously decreases. In this work, we focus on the regime where $D>D_c$.

For $K=0$, the structure reduces to a chiral helimagnetic state whose ground state is given by~\cite{kishine_coherent_2012},
\begin{equation}
    \theta_0(x)=\frac{\pi}{2}\,, \quad\quad\phi_0(x)=Q_0x\,,
\end{equation}
where $Q_0=D/2A$. As $K$ increases, a stronger DMI is required for the emergence of the CSS. In this paper, we study the case of $K>0$, where the ground state is the CSS, a type of chiral magnetic soliton lattice.


\section{Excitations}
\label{sec:sw}
\begin{figure}[t]
    \centering
    \includegraphics[width=0.45\textwidth]{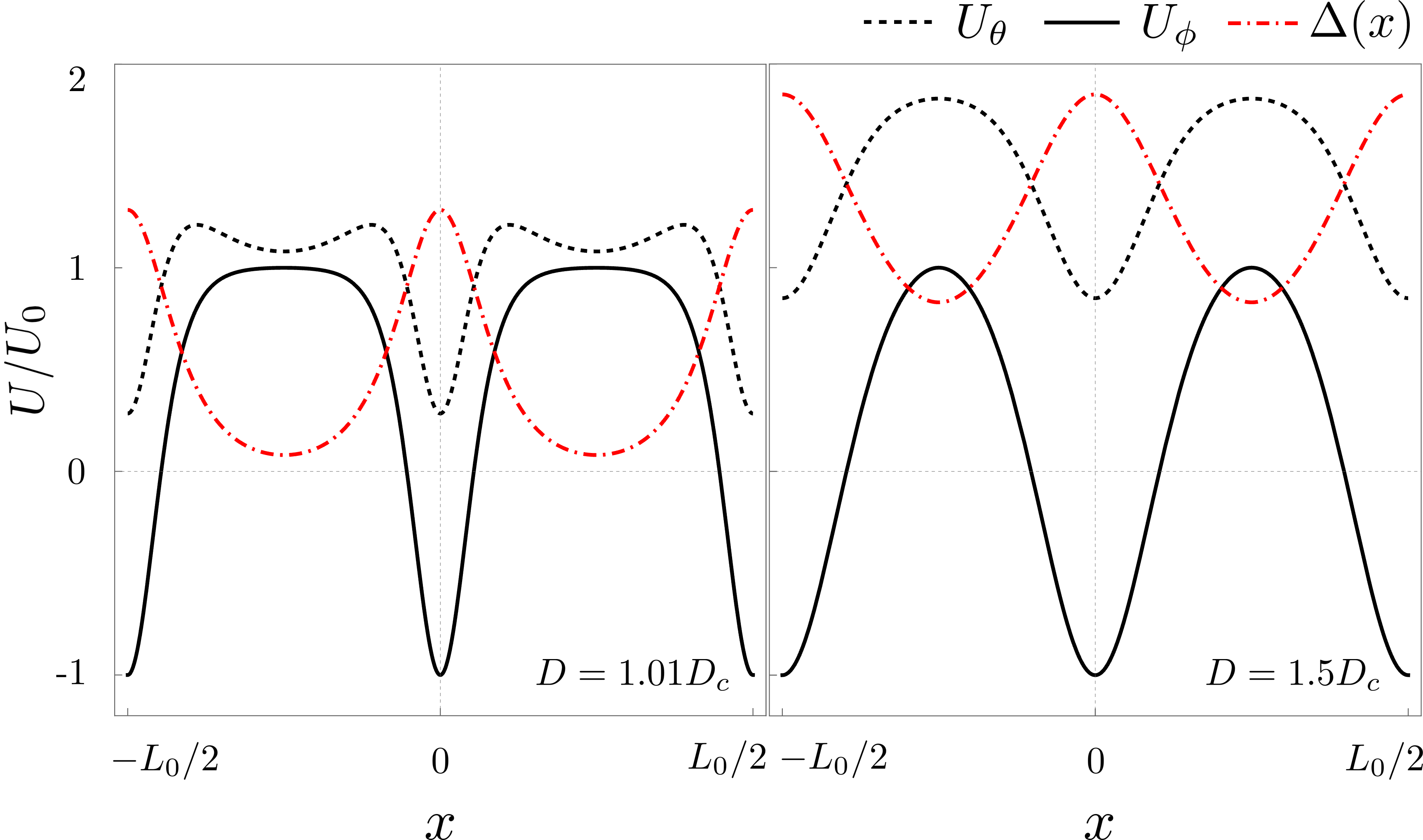} 
    \caption{Potentials of two operators, $\hat{H}_\theta$ [Eq.~(\ref{H_theta})] and $\hat{H}_\phi$ [Eq.~(\ref{H_phi})], and the difference between them, $\Delta(x)$ [Eq.~(\ref{delta})], for one period, where $U_0=1/\lambda^2$. The left panel shows the potential when $D=1.01D_c$, and the right panel illustrates it when $D=1.5D_c$.}
    \label{potential}
\end{figure}

As a preliminary to the discussion of the dynamics, we investigate the excitations of the ground state to construct an appropriate ansatz. Considering small deviations, $\delta\theta$ and $\delta\phi$, from the ground state of the two angle fields, the Hamiltonian density expanded up to the second order in the deviations is given by 
\begin{equation}
    \mathcal{H}[\theta_0+\delta\theta, \phi_0+\delta\phi]\simeq\mathcal{H}_0+\mathcal{H}_1+\mathcal{H}_2\,,
\end{equation}
where $\mathcal{H}_0=\mathcal{H}[\theta_0,\phi_0]$ and $\mathcal{H}_1$ is a surface term, which does not contribute to the dynamics. $\mathcal{H}_2$ is given by
\begin{equation}
    \mathcal{H}_2=A[\delta\theta\hat{H}_{\theta}\delta\theta+\delta\phi\hat{H}_{\phi}\delta\phi]\,,  
\end{equation}
which results in two differential equations governing the deviations. The two operators, $\hat{H}_{\theta}$ and $\hat{H}_{\phi}$, are given by~\cite{kishine2015theory}

\begin{gather}
     \hat{H}_{\theta}=-\frac{d^2}{dx^2}+\frac{1}{\lambda^2}(2\operatorname{sn}^2{\big(\frac{x}{\lambda\kappa},\kappa\big)}-1)+\Delta(x)\,,
     \label{H_theta}
     \\
     \hat{H}_{\phi}=-\frac{d^2}{dx^2}+\frac{1}{\lambda^2}(2\operatorname{sn}^2\big(\frac{x}{\lambda\kappa},\kappa\big)-1)
     \label{H_phi}
\end{gather}
with 
\begin{equation}
     \Delta(x)=\frac{D}{A\lambda\kappa}\operatorname{dn}\big(\frac{x}{\lambda\kappa},\kappa\big)-\frac{\kappa'^2}{\lambda^2\kappa^2}\,.\label{delta}
\end{equation}
Here, dn and sn are Jacobi functions and $\kappa'=\sqrt{1-\kappa^2}$. The two operators have the same form as the Schrödinger equation, whose potentials are shown in Fig.~\ref{potential}. For low $D$ shown in Fig.~\ref{potential}(a), $U_\phi$ can be interpreted as an array of P\"oschl-Teller potentials~\cite{poschl1933bemerkungen,kishine2015theory} and the difference of the two potentials, $\Delta(x)$, deforms the shape of $U_\theta$ from $U_\phi$ and shifts it upward, increasing the eigenenergies of modes in $\hat{H}_\theta$ than in $\hat{H}_\phi$. In Fig.~\ref{potential}(b), as $D$ increases, $U_\phi$ reduces to a sinusoidal function and $U_\theta$ increases. Therefore, $\Delta(x)$, which is proportional to $D$, behaves as a hindrance to mode excitations in $\hat{H}_\theta$ by lifting the potential.

\begin{figure*}[t]
    \centering
    \includegraphics[width=1\textwidth]{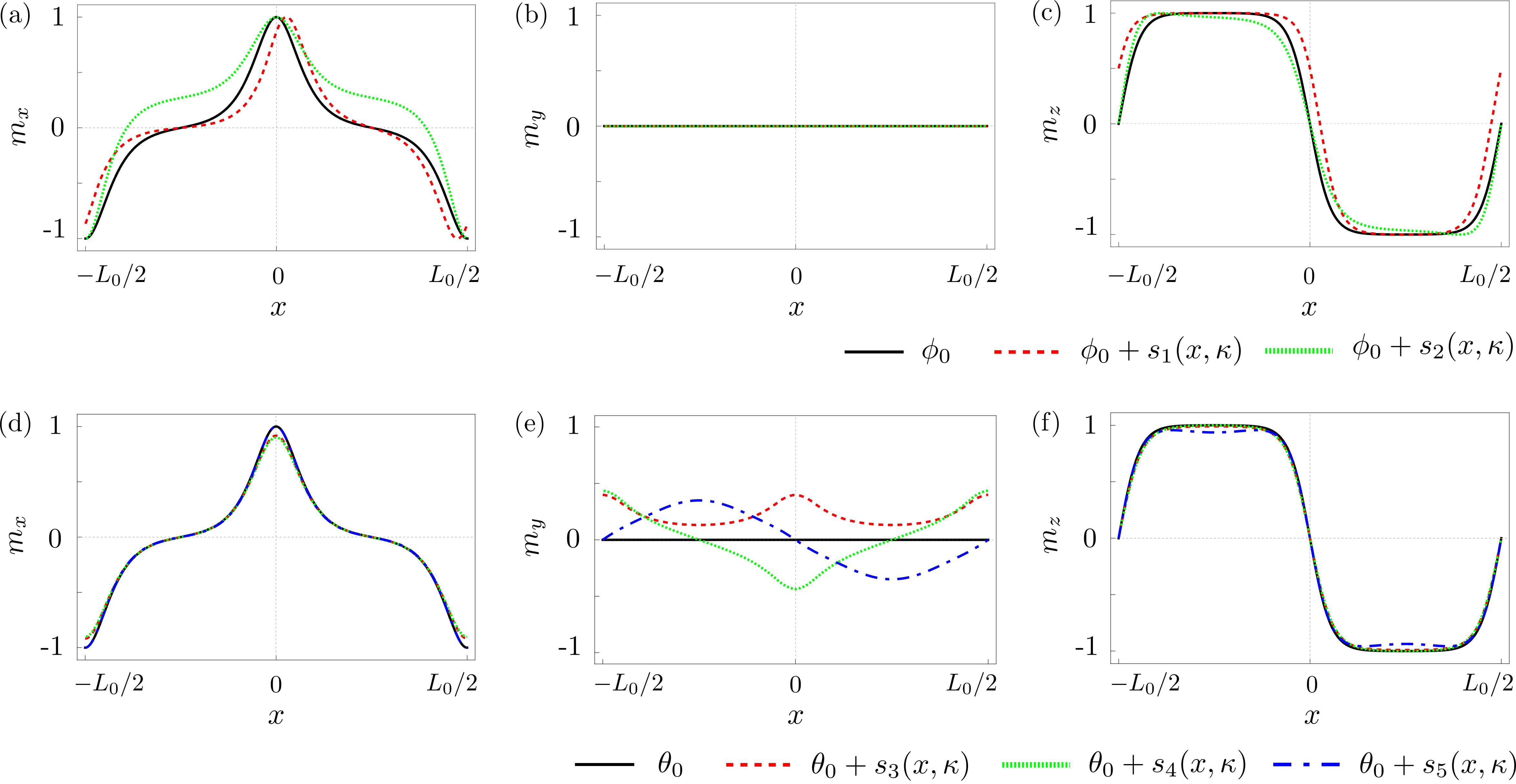}
    \caption{Plots showing how the modes used in the theory modify the ground state in components of the magnetization unit vector. $s_{i}(x,\kappa)$ is a function proportional to $u_{i}(s,\kappa)$ [Eqs. (\ref{ansatz1}) and (\ref{ansatz2})] for $D=2.858\,\mathrm{mJ/m^2}$.}
    \label{fig:modes}
\end{figure*}

The exact eigenfunctions for $\hat{H}_{\phi}$ are known to satisfy the Lam\'{e} equation, and the mathematical properties of the functions are rigorously studied in Ref.~\cite{kishine2015theory}. The eigenfunction, $\psi_{\bar{\sigma}}(s)$, introduced in Appendix~\ref{appb} is a periodic function. Here, $\bar{\sigma}$ is a complex parameter associated with quasi-momentum and eigenenergy [See Eqs.~(\ref{quasi}) and (\ref{eigenenergy})], and can be represented by a real parameter, $\sigma$, in two branches, 

\begin{alignat}{2}
    &\mathrm{Acoustic}\quad &&: \quad \bar{\sigma}=-i\sigma+K(\kappa)-iK'(\kappa)\,,
    \label{acoustic} \\
    &\mathrm{Optical}\quad &&: \quad \bar{\sigma}=i\sigma-iK'(\kappa)\,,
    \label{optical}
\end{alignat}
where $s=x/\lambda\kappa$ and $-K'(\kappa)\leq\sigma<K'(\kappa)$ with $K'(\kappa)$ defined by $K'(\kappa)=K(\kappa')$. 

In the periodic potentials, based on Bloch's theorem, we define the quasi-momentum, $\bar{q}(\bar{\sigma})$ [See Eq.~(\ref{quasi})], and, by imposing the periodic boundary condition,
\begin{equation}
    \psi_{\bar{\sigma}}(x)=\psi_{\bar{\sigma}}(x+L_0)\,,
\end{equation}
we identify two modes in the acoustic branch and infinitely many in the optical branch. One special mode in the acoustic branch is the zero mode, $X(t)$, which reflects the spontaneously broken translational symmetry from the infinite degeneracy of the ground states. Otherwise, there are no exact solutions for $\hat{H}_{\theta}$ due to the additional term, $\Delta(x)$. Therefore, we find the modes of $\hat{H}_{\theta}$ numerically.


\section{Dynamics}
\label{sec:dynamics}

\subsection{Theory}

In principle, the gliding motion of the CSS can be exactly described by considering an infinite number of modes in the theory. However, solving a large system of coupled differential equations is intractable. Therefore, assuming a weak perturbation, we select a small number of modes for each angle field, $\phi(x)$ and $\theta(x)$, which play dominant roles in the dynamics.

\begin{figure}[t]
    \centering
    \includegraphics[width=0.95\linewidth]{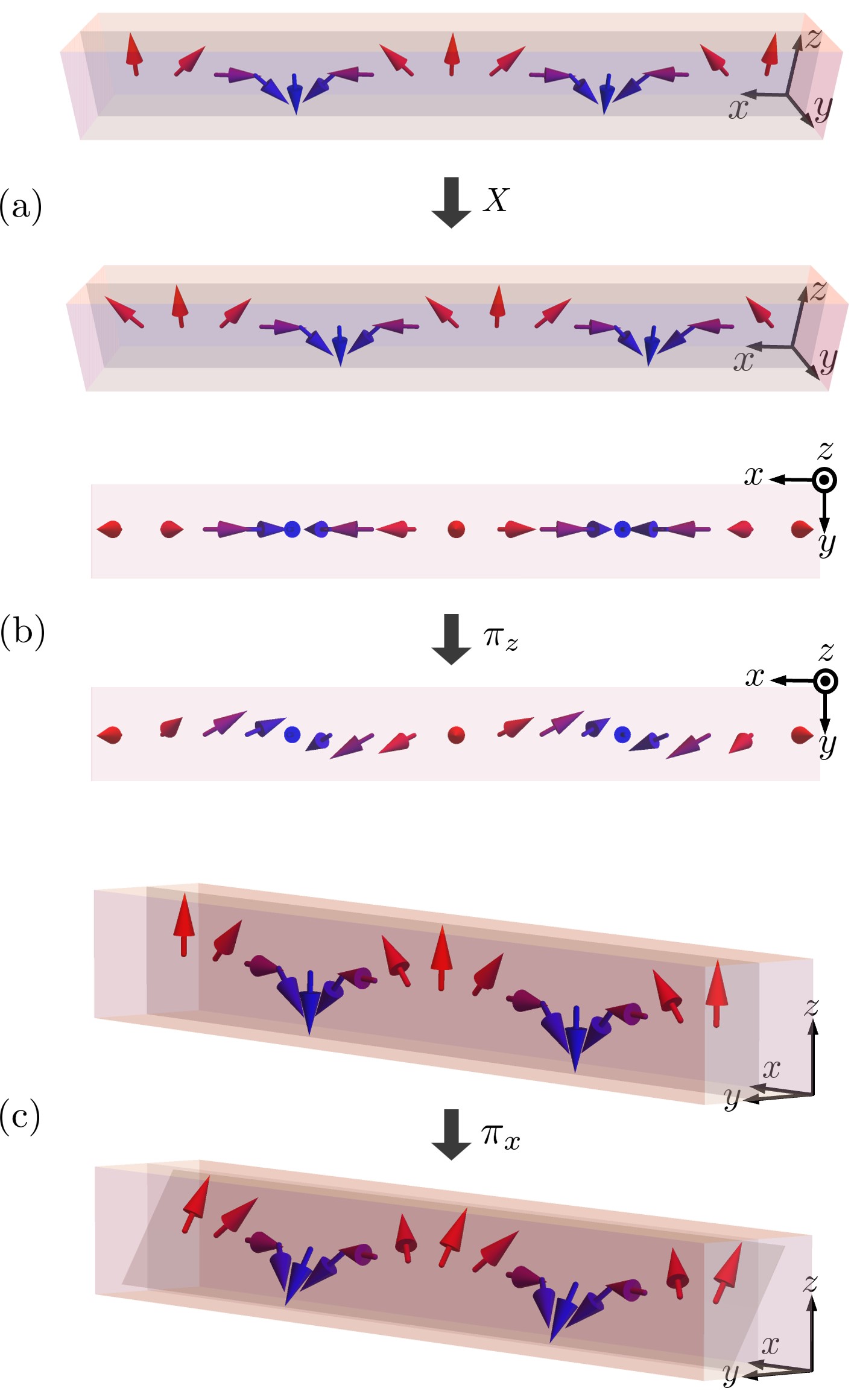} 
    \caption{Schematics explaining how the rotational modes operate on the CSS. Top (Middle) figure illustrates how $X$ ($\pi_z$) rotates the CSS about the $y$\,($z$) -axis. Bottom figure shows the rotation about the $x$-axis by $\pi_x$.}
    \label{fig.modeexplain}
\end{figure}

For each angle field, the three lowest-eigenenergy modes are used in the ansatz, which is given by 

\begin{align}
    \phi(x,t)=\,&\phi_0[x-X(t)]+\eta_z(t)u_1\big[\frac{x-X(t)}{\lambda\kappa},\kappa\big]\label{ansatz1}\\
    &+\eta_x(t)u_2\big[\frac{x-X(t)}{\lambda\kappa},\kappa\big]\nonumber\,,
\end{align}
\begin{align}
    \theta(x,t)&=\theta_0[x-X(t)]+\eta_y(t)u_3\big[\frac{x-X(t)}{\lambda\kappa},\kappa\big]\label{ansatz2}\\
    &+\pi_z(t)u_4\big[\frac{x-X(t)}{\lambda\kappa},\kappa\big]+\pi_x(t)u_5\big[\frac{x-X(t)}{\lambda\kappa},\kappa\big]\nonumber\,.
\end{align}
Here, $u_1$ and $u_2$ ($u_3$, $u_4$ and $u_5$), are excitation modes of $\hat{H}_\phi$ ($\hat{H}_\theta$) with subscripts reflecting increasing eigenenergy, respectively, and $X(t)$ is zero mode from $\hat{H}_\phi$. Here, $\eta_i(t)$ and $\pi_i(t)$ are dynamical variables of these modes. Figure \ref{fig:modes} shows how those modes induce deviations from the ground state where solitons are located at $x=0,\,\pm L_0/2$. The mode, $u_1$, shifts the positions of solitons with $m_x=1\,(-1)$ toward positive~(negative) $x$-direction, resulting in translational motion of solitons. In contrast, $u_2$, instead of moving solitons, expands~(shrinks) their width at $x=0\,(\pm L/2)$. Both modes do not affect $m_y$. On the other hand, $u_3$\,($u_4$) increases $m_y$ at the soliton positions in the same~(opposite) direction, while $u_5$ enhances $m_y$ on domains in the opposite direction. These modes have a negligible impact on $m_x$ and $m_z$. From a more physical perspective, $u_1$, $u_2$ and $u_3$ increase $m_z$, $m_x$ and $m_y$, respectively, whereas $X$, $u_4$ and $u_5$ correspond to rotation of the CSS about the $y$-, $z$- and $x$-axis, respectively [See Fig. \ref{fig.modeexplain}]. 

The Lagrangian density is given by
\begin{equation}
    \mathcal{L}[\theta(x,t), \phi(x,t)]=\mathcal{K}-\mathcal{H}=\mathcal{J}(\cos{\theta}-1)\dot{\phi}-\mathcal{H}\,,
    \label{lagrangian}
\end{equation}
where the first term, $\mathcal{K}$, represents the Berry phase of precessing spins~\cite{PhysRevLett.57.1488}, $\mathcal{J}=M/\gamma$ is the spin density and $\gamma$ is the gyromagnetic ratio of free electrons. After integration over one period, the Lagrangian expanded up to the second order in dynamical variables is given by


\begin{align}
    L=\,&G_y \eta_y\dot{X}+G_z\pi_z\dot{\eta}_z+G_x\pi_x\dot{\eta}_x\label{lagrangian2}
    \\
    &-\sum_{i}C_i\eta^2_i-C_z'\pi^2_z-C_x'\pi^2_x-\sum_{i}Q_i\bar{H}_i\eta_i \nonumber\,,
\end{align}
where $\mathbf{H}=(\bar{H}_x,\bar{H}_y,\bar{H}_z)=(H_x,H_y,H_z)\sin{\omega t}$, $i\in\{x,y,z\}$ and omit the $(t)$ notation. Here, $\omega$ is the angular frequency of the AC magnetic field. The coefficients are determined by the integration of space-dependent functions composed of complex Jacobi functions and excitation modes [See Appendix~\ref{appc}]. The gyroscopic coefficients, $G_x$, $G_y$, and $G_z$, couple ($\eta_x$, $\pi_x$), ($\eta_y$, $X$) and ($\eta_z$, $\pi_z$), respectively. They arise from the Berry phase term and are proportional to the spin density, $\mathcal{J}$. The coefficients, $C_z$, $C_x$, $C_y$, $C_z'$ and $C_x'$, act as spring constants. Each is proportional to the eigenenergy of its corresponding mode, $\eta_z$, $\eta_x$, $\eta_y$, $\pi_z$ and $\pi_x$, which depends on the system parameters in the Hamiltonian density such as $A$, $K$, and $D$. (Note that, although $\eta_i$ and $\pi_i$ are dynamical variables of modes, \{$u_i$\}, we hereafter refer to them simply as modes by assigning the physical meanings discussed above for brevity.) The coefficient, $Q_i$, is the coupling constant between the AC magnetic fields and the modes, which is proportional to the magnetization, $M$. In the same manner, we expand the Rayleigh dissipation function up to the third order in modes: 

\begin{align}
    R=\,&D_X \dot{X}^2+\sum_{i}D_i\dot{\eta}_i^2+{D_z'}\dot{\pi}_z^2+{D_x'}\dot{\pi}_x^2\\&\nonumber +[D_{xz}\eta_x\dot{X}\dot{\eta}_z+D_{xz}'\pi_x\dot{X}\dot{\pi}_z+(x\leftrightarrow z)]\,,
\end{align}
where the coefficients are given in Appendix \ref{appc}.
From the Lagrangian and the Rayleigh dissipation function, equations of motion (EOMs) for each mode are derived, and the EOM for $X$ is given by


\begin{equation}
    2D_X\dot{X}+[D_{xz}\eta_x\dot{\eta}_z+D_{xz}'\pi_x\dot{\pi}_z+(x\leftrightarrow z)]+G_y\dot{\eta}_y=0\,.\label{eom1}
\end{equation}

Here, we provide a few remarks. In Eq.~(\ref{lagrangian2}), $\eta_z$, $\eta_x$, and $\eta_y$ are coupled with $H_z$, $H_x$, and $H_y$, respectively. There are three canonically conjugate pairs: ($X$, $\eta_y$), ($\eta_z$, $\pi_z$) and ($\eta_x$, $\pi_x$). The mode, $\eta_y$, serves as a source of the velocity, $\dot{X}$, except for the dissipative channels, but it does not contribute to the average velocity due to the time-odd nature in the EOM. To capture the dominant contribution to the average velocity, we include inter-mode coupling terms in the dissipative channels at the lowest order. Therefore, we expand the Rayleigh dissipation function up to the third order. The coupling terms involve $\eta_z$ and $\eta_x$, which play the roles of a translational motion and a variation of width, respectively, as well as $\pi_z$ and $\pi_x$, which describe the out-of-plane fluctuations of spins at the soliton cores and on the domains, respectively.

To go further, we focus on the linear response to an AC magnetic field with an angular frequency, $\omega$, by assuming $\eta_i(t)=\tilde{\eta}_i\sin{(\omega t-\delta_i)}$ and $\pi_i(t)=\tilde{\pi}_i\sin{(\omega t-\delta_i')}$, and substitute them into the EOMs. After time-averaging Eq.~(\ref{eom1}) over one period, we obtain an expression for the average velocity as a function of the angular frequency, $\omega$, and the AC magnetic field, $\mathbf{H}$: 
\begin{equation}
    \bar{V}=\frac{\omega}{2D_X}[D_{zx}\tilde{\eta}_z\tilde{\eta}_x\sin{(\delta_z-\delta_x)}+D_{zx}'\tilde{\pi}_z\tilde{\pi}_x\sin{(\delta_z'-\delta_x')}]
    \label{avgvel}
\end{equation}
with $\delta_i$, $\delta_i'$, $\tilde{\eta}_i$ and $\tilde{\pi}_i$, which are also given as functions of $\omega$ and $\mathbf{H}$ in Appendix~\ref{appc}.
This is our first main result: the expression for the average velocity of the CSS, $\bar{V}$, driven by an AC magnetic field.

\begin{figure*}[t]
    \centering
    \includegraphics[width=1\linewidth]{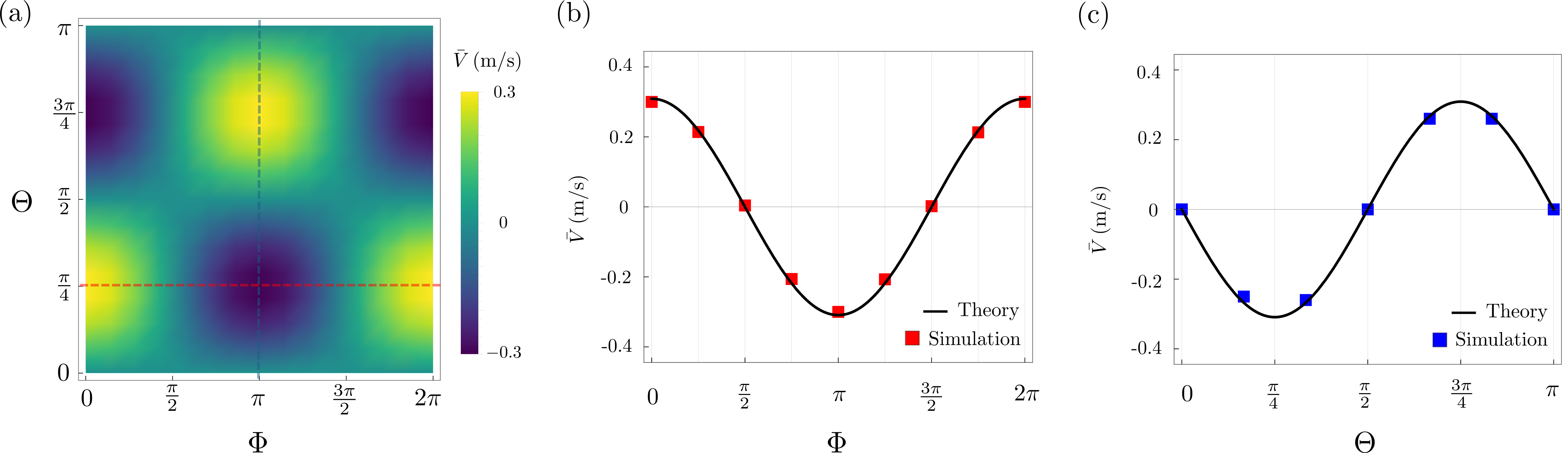} 
    \caption{The plots of the average velocity, $\bar{V}$ [Eq.~(\ref{avgvel})], for (a) all solid angles from the theory and the comparison with simulations for (b) red dashed line~($\Theta=\pi/4$) and (c) blue dashed line~($\Phi=\pi$), where $D=2.858\,\mathrm{mJ/m^2}$, $f=10\,\mathrm{GHz}$ and $H=30\,\mathrm{mT}$.}
    \label{fig.solid angle}
\end{figure*}

Since many of our results pertain to the expression, let us make several observations. First, the average velocity requires finite dissipation, because the coefficients, $D_i$, all vanish when $\alpha=0$. 
In the first term which is proportional to $D_{zx}$, $\eta_z$ and $\eta_x$ are coupled through the dissipation and contribute to the average velocity. As previously discussed their roles [See Fig.~\ref{fig:modes}], it is analogous to the mechanism of the DW for the unidirectional gliding motion under an AC magnetic field~\cite{moon2017domain}. The other source for the average velocity involves the modes, $\pi_z$ and $\pi_x$, which were not considered in the previous DW study. This difference leads us to conjecture that an additional collective coordinate may be required to more accurately describe the DW dynamics under an AC magnetic field. In the spirit of the collective coordinate approach, additional degrees of freedom can be introduced to capture more complex dynamics at the cost of mathematical simplicity. For example, an extended theory using four coordinates was proposed, in which an additional coordinate, $\chi$, representing a geometric tilt angle of DW, was introduced beyond the conventional three collective coordinates, $X$, $\Phi$ and $\Delta$~\cite{nasseri2018collective,RevModPhys.95.015003}. The mode was introduced to explain the tilting of fast-moving DWs in perpendicularly magnetized thin films with DMI~\cite{ryu2012current,PhysRevLett.111.217203}. Furthermore, in the last section of Ref.~\cite{moon2017domain}, the authors pointed out that the deviation between simulations and the theory originates from the oscillation of domains which was not accounted for in their model. In our theory, such effects are captured by $\pi_x$, which rotates the domains about the $x$-axis. The inclusion of the mode completes the set of canonically conjugate pairs, and, as shown in Fig. \ref{fig.modeexplain}, reveals three rotational modes for each axis.

Secondly, from the solutions of $\tilde{\eta}_{x}$, $\tilde{\pi}_{x}$, $\tilde{\eta}_{z}$ and $\tilde{\pi}_{z}$, we identify that the expression for average velocity is proportional to $H_x$ and $H_z$, and is independent of $H_y$ [See Appendix~\ref{appc}], i.e., $\bar{V}\propto H_x H_z$. This result differs from the case of a single N\'{e}el type DW~\cite{moon2017domain} and provides an example of the fundamentally different dynamics between the DW and the CSS. In the CSS (DW), the zero mode, $X$, is affected by $H_y$ through $\eta_y$ ($\Phi$), which corresponds to an increase in $m_y$ (a rotation about the $z$-axis). The modes, $\eta_y$ and $\Phi$, respond to $H_y$ in different ways.

Thirdly, the average velocity shows a maximum velocity at a specific driving frequency. This behavior is consistent with our predictions that under a DC magnetic field~($\omega=0$) the spin structure is saturated to a new ground state and under a high frequency AC magnetic field the average velocity is suppressed due to failure to satisfy the resonance condition. However, our theory has a limitation rooted in the assumption that $\eta_i(t)$ and $\pi_i(t)$ are sinusoidal functions. We expect that our theory remains valid in the sufficiently high frequency regime where they follow the sinusoidal form.




\subsection{Results and Discussion}

In this section, we compare our theory with simulation results. For micromagnetic simulations, we use MuMax3, which is an open-source software solving Landau-Lifshitz-Gilbert~(LLG) equation~\cite{vansteenkiste2014design}. The constant parameters used in simulations are $A=10^{-11}\,\mathrm{J/m}$, $K=5\times10^5\,\mathrm{J/m^3}$, $M=1000\,\mathrm{kA/m}$, $\lambda=\sqrt{A/K}=4.47\,\mathrm{nm}$, a damping factor $\alpha=0.1$ and a gyromagnetic ratio $\abs{\gamma}=1.76\times10^{11}\,\mathrm{rad/s\cdot T}$~\cite{moon2017domain}. The varying parameters are the DMI constant, $D$, the amplitude of the external AC magnetic field, $H$, and its frequency, $f$. For simplicity, we neglect the effect of demagnetization on both simulations and the theory in the main text, and investigate its dynamical effect in Appendix \ref{appc}. To extract the average velocity, $\bar{V}$, from simulations, we track a specific soliton, e.g., the position where $m_z=0$, in the spin structure, and the trajectory is fitted over several periods which depends on the frequency of the AC magnetic field.

\begin{figure}[t]
    \centering
    \includegraphics[width=0.8\linewidth]{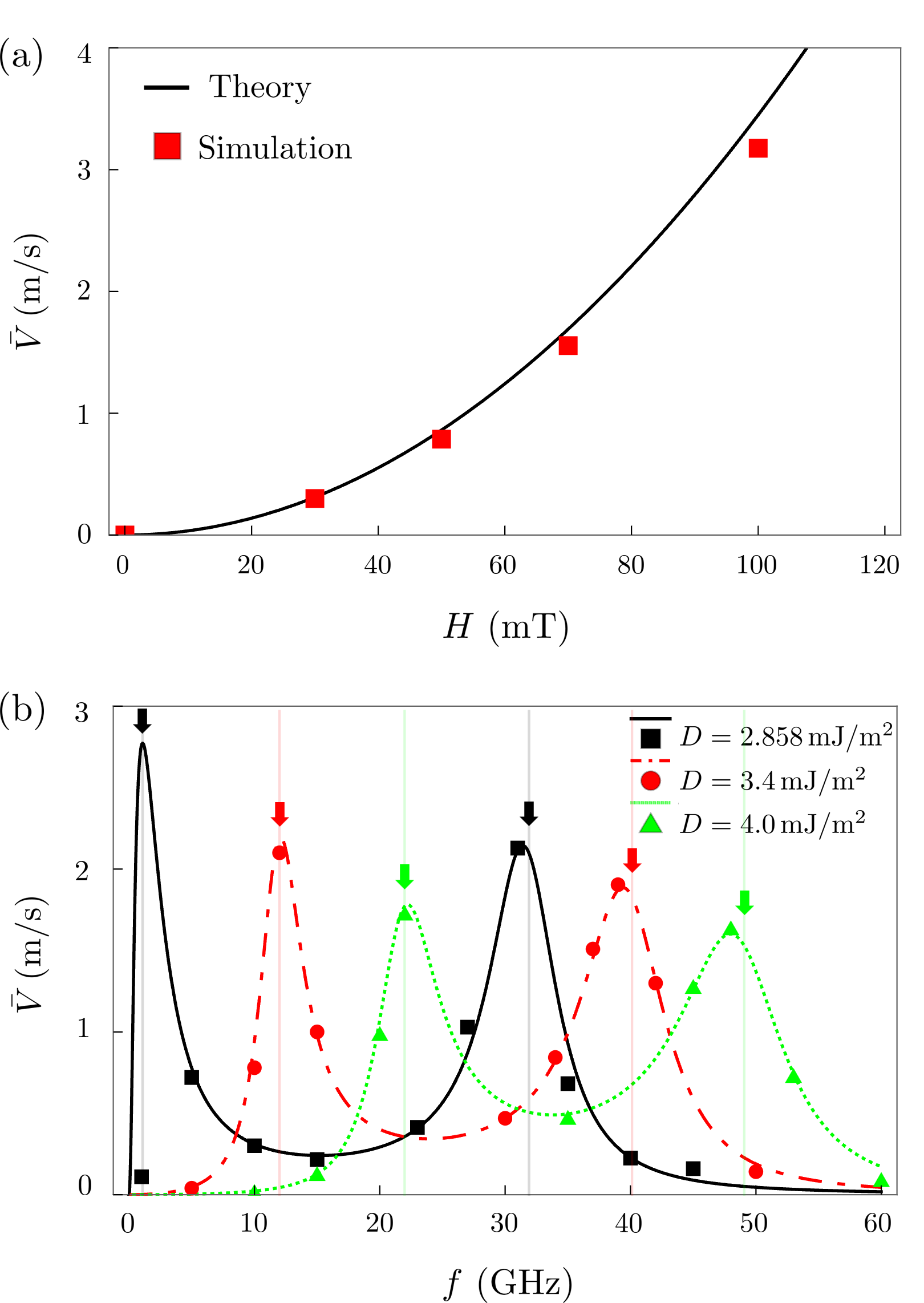} 
    \caption{The plots of the average velocity, $\bar{V}$ [Eq.~(\ref{avgvel})], at a fixed solid angle, $(\Theta,\Phi)=(\pi/4,0)$. (a) $\bar{V}$-$H$ plot where $D=2.858\,\mathrm{mJ/m^2}$ and $f=10\,\mathrm{GHz}$. (b) $\bar{V}$-$f$ plots for different $D$, where $H=30\,\mathrm{mT}$. The vertical lines and arrows indicate the natural frequencies, $\omega_1/2\pi$ and $\omega_2/2\pi$ [Eqs. (\ref{resonance}) and (\ref{resonance2})], which are essentially identical to the resonance frequencies for each $D$.}
    \label{fig.fieldfreq}
\end{figure}

Figure \ref{fig.solid angle}(a) shows the theoretical result of $\bar{V}$ as a function of solid angle. From Eq.~(\ref{avgvel}), we derive an equation of $\bar{V}$ in terms of the solid angle, $\Theta$ and $\Phi$: 
\begin{equation}
    \bar{V}\propto H_x H_z\propto H^2 \sin{2\Theta}\cos{\Phi}\,,
    \label{anglevel}
\end{equation}
where 
\begin{equation}
    (H_x,H_y,H_z)=H(\sin{\Theta}\cos{\Phi},\sin{\Theta}\sin{\Phi},\cos{\Theta})\,.
\end{equation}
Here, $\bar{V}$ vanishes where $\Phi=\pi/2,\,3\pi/2$ and $\Theta=0,\,\pi/2,\,\pi$, i.e., when either $H_x$ or $H_z$ is zero. Figure \ref{fig.solid angle}(b) and Figure \ref{fig.solid angle}(c) present simulation results along the two dashed lines in Fig.~\ref{fig.solid angle}(a), exhibiting good agreement with our theoretical predictions. Figure \ref{fig.fieldfreq}(a) shows the dependence of $\bar{V}$ on the intensity of the magnetic field, $H$, and the simulation results exhibit the same quadratic behavior as expected by Eq.~(\ref{anglevel}). Next, we investigate the frequency dependence of $\bar{V}$. Figure \ref{fig.fieldfreq}(b) shows the results for three different $D$. In each case, two resonance angular frequencies, $\bar{\omega}_1$ and $\bar{\omega}_2$, exist at which $\bar{V}$ reaches local maxima.

The existence of the two resonance frequencies for the average velocity can be understood in the following way. In Eq.~(\ref{avgvel}), we identify that the two pairs of the modes, $(\eta_{z},\,\pi_{z})$ and $(\eta_{x},\,\pi_{x})$, contribute to the average velocity. To elucidate the origin of the resonance angular frequencies, $\bar{\omega}_1$ and $\bar{\omega}_2$, we analyze those modes in the absence of the dissipation, which provides clear insights into the underlying dynamics. The EOMs for the first pair are given by
\begin{alignat}{2}
    &\eta_z:\quad&&\, Q_z \bar{H}_z+2C_z\eta_z+G_z\dot{\pi}_z=0\,,\\
    &\pi_z:\quad&&\, G_z\dot{\eta}_z-2C_z'\pi_z=0\,.
\end{alignat}
These EOMs without the external magnetic field lead us to two coupled harmonic oscillators with phase difference of $\pi/2$ and amplitudes differing by a factor of $\sqrt{C_z'/C_z}$. Reintroducing the external field and the damping, the mode coupled with the external field follows the dynamics of the forced harmonic oscillator. It is well known that the oscillating coordinate reaches its maximum amplitude and has a $\pi/2$ phase difference with the driving force when the frequency of the driving force is equal to the natural frequency of the system. As the driving frequency deviates from the natural frequency, the amplitude decreases and the phase difference gradually deviates from $\pi/2$, while $\pi_z$ follows $\eta_{z}$ with a nearly constant phase offset of $\pi/2$. The natural angular frequency of the subsystem is given by 
\begin{equation}
    \omega_1=\sqrt{\frac{4C_zC_z'}{G_z^2}}\propto\sqrt{\epsilon_1\epsilon_4}\,,\label{resonance}
\end{equation}
where $\epsilon_i$ is an eigenenergy of the excitation mode, $u_i$. In the same manner, from $\eta_x$ and $\pi_x$, the second natural angular frequency, $\omega_2$, which exceeds $\omega_1$, is given by
\begin{equation}
    \omega_2=\sqrt{\frac{4C_xC_x'}{G_x^2}}\propto\sqrt{\epsilon_2\epsilon_5}\,.\label{resonance2}
\end{equation}

The natural angular frequencies, $\omega_1$ and $\omega_2$, of the two subsystems are identical to the resonance angular frequencies, $\bar{\omega}_1$ and $\bar{\omega}_2$, at which the average velocity, $\bar{V}$ [Eq.~(\ref{avgvel})], is locally maximized [See Fig. \ref{fig.fieldfreq}(b)]. This agreement arises because $\bar{V}$ is maximized when the amplitudes of the modes, $\tilde{\eta}_i$ and $\tilde{\pi}_i$, are enhanced and the phase difference between the two coupled modes approaches $\pi/2$. At the natural angular frequency, $\omega_{1(2)}$, $\tilde{\eta}_{z(x)}$ and $\tilde{\pi}_{z(x)}$ reach their maxima, and the phase differences, $\delta_{z}-\delta_{x}$ and $\delta_{z}'-\delta_{x}'$, approach $\pi/2$ for both natural angular frequencies. The separation of the two resonance angular frequencies guarantees this behavior, allowing each resonance frequency to selectively induce a $\pi/2$ phase shift in its corresponding mode while deviating it from $\pi/2$ in the mode involved with the other resonance.
Additionally, we show the DMI dependence of $\bar{V}$ at a fixed solid angle and frequency in Fig.~\ref{fig.dmi}(a). As mentioned in Fig.~\ref{fig.fieldfreq}, there is a specific value of $D$ at which the average velocity is maximized where $\bar{\omega}_1/2\pi=10\,\mathrm{GHz}$. As $D$ increases, $\bar{V}$ is suppressed due to the increase in the resonance frequency, which is a behavior distinct from that observed in DWs~\cite{moon2017domain}. Numerical confirmation of the average velocity of the CSS [Eq.~(\ref{avgvel})] and the explanation of the two resonance frequencies for the average velocity are our second main result.





\subsection{Spin motive force}

\begin{figure}[t]
    \centering
    \includegraphics[width=0.8\linewidth]{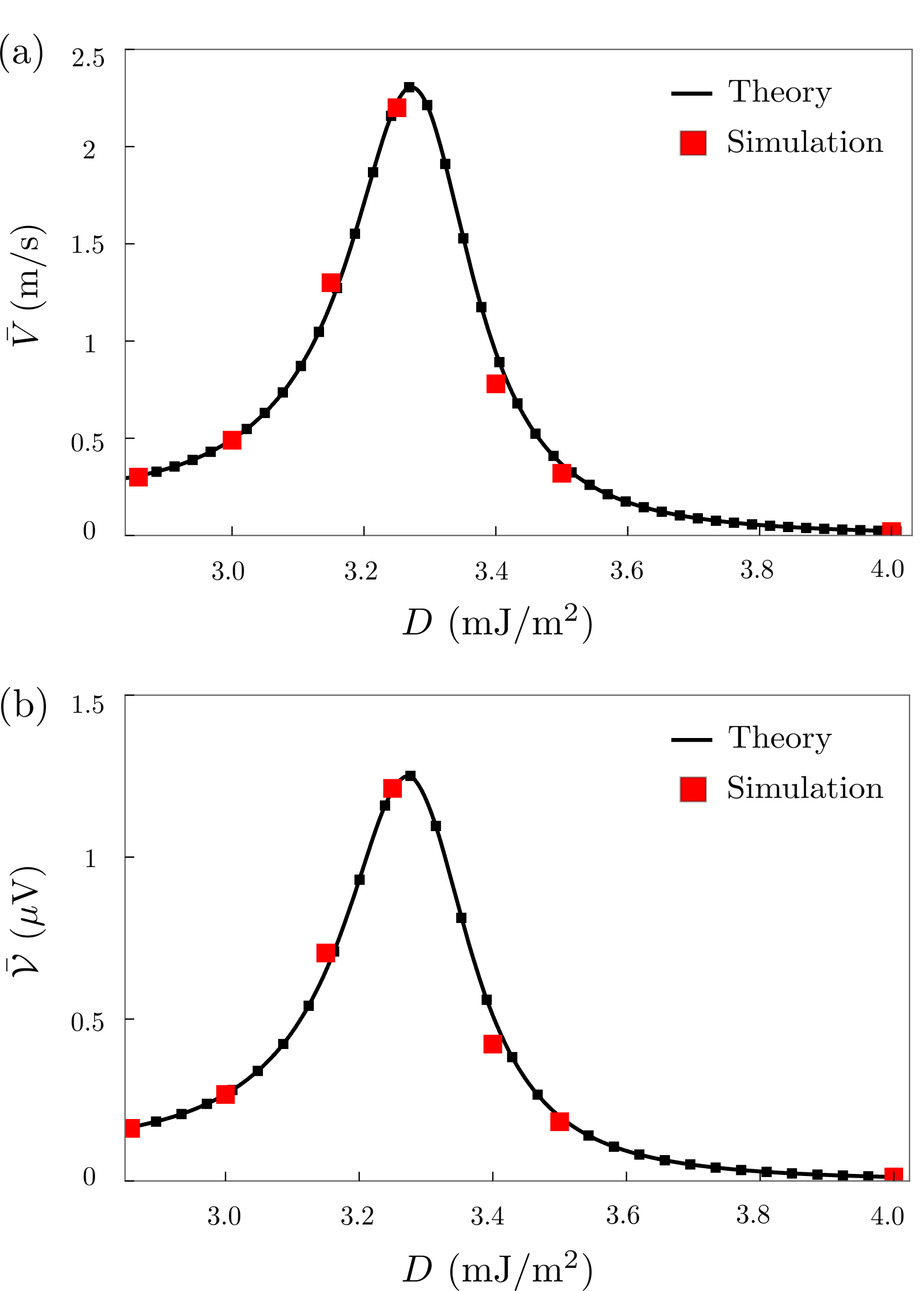} 
    \caption{The plots of the average velocity, $\bar{V}$ [Eq.~(\ref{avgvel})], and the DC voltage, $\bar{\mathcal{V}}$ [Eq.~(\ref{avgvoltage})], at a fixed solid angle, $(\Theta,\Phi)=(\pi/4,0)$. (a) $\bar{V}$-$D$ plot and (b) $\bar{\mathcal{V}}$-$D$ plot for a spin-$\uparrow$ electron where $f=10\,\mathrm{GHz}$ and $H=30\,\mathrm{mT}$. The black solid lines are obtained by interpolating the discrete small black markers. }
    \label{fig.dmi}
\end{figure}

In the preceding sections, we demonstrated that the CSS exhibits a unidirectional gliding motion under an AC magnetic field. Since this process involves a dynamical deformation of the spin structure, the emergence of a DC voltage is expected. 
Here, we evaluate SMF originating from the Rashba effect. In a ferromagnetic heterostructure identical to the system considered in our theory, spin-orbit coupling induced by the broken inversion symmetry at the interface can lead to spin-polarized currents, thereby exerting finite torques on the local magnetization, which is called spin-orbit torque. Recalling the reciprocal relation, the magnetization dynamics under the Rashba interaction can give rise to SMF. The force acting on conduction electrons is described by
\begin{equation}
    \mathbf{F}^{R}_\pm=\pm\alpha_R\frac{m_e}{\hbar}[(\hat{\mathbf{z}}\times\partial_t\mathbf{m})+\beta\hat{\mathbf{z}}\times(\mathbf{m}\times\partial_t\mathbf{m})]\,,
\end{equation} 
where $\pm$ applies to the spin-$\uparrow$~($\downarrow$) electrons, $\alpha_R$ denotes Rashba constant and $m_e$ is the effective mass of conduction electrons~\cite{PhysRevB.87.054403,PhysRevLett.108.217202}. Note that the spin-$\uparrow$ ($\downarrow$) electrons are antiparallel (parallel) to local magnetization, $\mathbf{m}$. We find that the non-adiabatic contribution to the DC voltage for one period, $L_0$, attributed to the Rashba effect, is substantial~\cite{myfootnote}. The expression of the DC voltage is given by
\begin{equation}
    \bar{\mathcal{V}}_\pm\simeq\mp\frac{2\pi\alpha_R \beta m_e}{e\hbar} \bar{V}\,,
    \label{avgvoltage}
\end{equation}
where $\bar{V}$ is the average velocity given in Eq.~(\ref{avgvel}) and $\beta$ is a constant parameterizing the intensity of the non-adiabatic contribution~\cite{PhysRevB.77.134407}. Here, $\beta$ is identical to that appears in non-adiabatic spin-transfer torque based on the Onsager reciprocity principle~\cite{landau2013course}. In this paper, we set $\beta=\alpha=0.1$~\cite{pollard2012direct,PhysRevB.95.184419}. Figure \ref{fig.dmi}(b) shows the theoretical prediction of the DC voltage arising from the Rashba effect. For the calculation, we use parameters for $\mathrm{Pt/Co(0.6\,nm)/AlO_x}$, which exhibits a comparable DMI constant and a large Rashba constant \cite{PhysRevLett.108.217202,PhysRevB.91.180405}, and assume that the effective mass is the mass of a free electron. The resulting DC voltage reaches a maximum on the order of microvolts.


At this point, we highlight several remarks. First, the expression for $\bar{\mathcal{V}}$ contains $\bar{V}$, indicating that the DC voltage also requires finite dissipation. Furthermore, the CSS provides a significant advantage in amplifying the voltage by a factor of $N$, the number of solitons in the system, i.e., the total voltage is given by $\bar{\mathcal{V}}_{total}=N\bar{\mathcal{V}}$. This advantage opens up the possibility for using the CSS as an effective magnetic rectifier, capable of converting AC magnetic fields into DC voltages, which may be exploited for energy-harvesting devices~\cite{sharma2024nanoscale}. However, several challenges should be overcome for practical applications, such as the requirement for a large $\beta$, the still small magnitude of the DC voltage, and the difficulty of fabricating samples with strong DMI. This is the last of our main results: the theoretical demonstration of a finite DC voltage from the SMF induced by the Rashba effect [Eq.~(\ref{avgvoltage})] under an AC magnetic field, which holds promise as an energy-harvesting mechanism.



\section{Conclusion}
\label{sec:conclusion}

In this paper, we studied the dynamics of the CSS, a noncollinear nonlinear spin structure induced by strong interfacial DMI, under an AC magnetic field. We first suggested the simplest model Hamiltonian describing the CSS in a ferromagnetic thin film with an easy-axis anisotropy and demonstrated the ground state and excitations of the system. 
We employed the Lagrangian formalism with a collective coordinate and excitation modes to investigate the dynamics. Under the assumption of a weak perturbation, we derived an equation for the average velocity and identified an analogy and a difference between the mechanism of the gliding motion in the CSS and that in the DWs. The equation is proportional to $H_x$ and $H_z$, while remaining insensitive to $H_y$, and exhibits a quadratic dependence on the magnitude of the AC magnetic field. The average velocity, $\bar{V}$ shows two resonance frequencies, and through the analysis of the modes associated with the average velocity, we identified the origin of the resonance frequencies. To confirm our theory, we performed the micromagnetic simulations and the results showed good agreement with our theory under different solid angles, intensity, frequency, and DMI. 
The observed gliding motion naturally led to an examination of SMF. We found that the DC voltage is attributed to the non-adiabatic process from the Rashba effect, and the evaluation of the DC voltage motivates further investigation into future applications.

As future work, several approaches can be explored to enhance the DC voltage along with the average velocity, $\bar{V}$. One direction is to construct a synthetic antiferromagnet, which supports higher DW velocities than conventional ferromagnets~\cite{yang2015domain,duine2018synthetic}. In addition, optimizing the system parameters may further improve performance. For example, tuning the easy-axis anisotropy constant or the DMI  through various experimental techniques~\cite{PhysRevB.79.024423,PhysRevApplied.13.044029} can yield a higher average velocity, even under substantially reduced magnetic fields~\cite{note2}.


\begin{acknowledgments}
D.H.H. and S.K.K. were supported by Brain Pool Plus Program through the National Research Foundation of Korea funded by the Ministry of Science and ICT (2020H1D3A2A03099291) and National Research Foundation of Korea(NRF) grant funded by the Korea government(MSIT) (2021R1C1C1006273). K.W.M. was supported by the Technology Innovation Program (RS-2022-00154785) funded by the Ministry of Trade, Industry \& Energy (MOTIE, Korea) and the Nano \& Material Technology Development Program through the National Research Foundation of Korea(NRF) funded by Ministry of Science and ICT(RS-2024-00451261). K.J.K was supported by the National Research Foundation of Korea (RS-2023-00275259).
\end{acknowledgments}

\appendix

\section{Ground State}
\label{appa}
\subsection{Sine-Gordon equation}

To find the ground state and its period under strong DMI and easy-axis anisotropy, we set $\theta_0(x)=\pi/2$, which reduces the Hamiltonian density to  
\begin{equation}
    \mathcal{H}[\mathbf{m};\theta_0(x)=\pi/2]=
        A{\phi'}^2
        +K\sin^2{\phi}
        -D\phi'\,.
\end{equation}
The functional derivative of the Hamiltonian density for $\phi$ yields an equation that takes the form of the sine-Gordon equation~\cite{barone1971theory}:
\begin{equation}
    \frac{\delta\mathcal{H}}{\delta \phi_0}=-2A\phi_0''+K\sin{2\phi_0}=0\,,
\end{equation}
which gives the ground state, 
\begin{equation}
    \phi_1(x)=2 \operatorname{am}\big(\frac{x}{\lambda\kappa},\kappa\big) \quad \Rightarrow \quad \phi_0(x)=\frac{\pi}{2}+\operatorname{am}\big(\frac{x}{\lambda\kappa},\kappa\big)\,,
\end{equation}
where $2\phi_0=\pi+\phi_1$ and $\lambda=\sqrt{A/K}$.

\subsection{Period}

The ground state, $\phi_0(x)$, is expressed in terms of the Jacobi amplitude function, $\operatorname{am}$, with a period given by 
\begin{equation}
    L_0=4 \lambda \kappa K(\kappa)\,, \label{defperiod}
\end{equation}
where $K(\kappa)$ is the complete elliptic integral of the first kind. The elliptic modulus, $\kappa$, determining the period is chosen to minimize an energy per soliton, $\mathcal{E}_{s}(\kappa)$:

\begin{equation}
    \mathcal{E}_{s}(\kappa)
    =\frac{1}{L_0}\int dx\,\mathcal{H}(x)=\frac{A}{\kappa^2\lambda^2}[\frac{2E(\kappa)}{K(\kappa)}-1]-\frac{2\pi D}{4\kappa K(\kappa)\lambda}\,,
\end{equation}
\begin{equation}
    \frac{d \mathcal{E}_{s}(\kappa)}{d\kappa}=0\quad\Rightarrow\quad \kappa=\frac{4AE(\kappa)}{\pi \lambda D}=\frac{4E(\kappa)}{\pi D}\sqrt{A K}\,,
\end{equation}
where $\partial_x \phi_0=q\operatorname{dn}(qx,\kappa)$, $q=4K(\kappa)/L_0=1/\lambda\kappa$ and $E(\kappa)$ is the elliptic integral of the second kind. However, because of the constraint, $0\leq \kappa \leq1$, for fixed values of $A$ and $K$, there exists a critical DMI, $D_c$, at which IC-C phase transition occurs. Figure \ref{fig:period} shows $D_c\simeq2.847\times10^{-3}\,\mathrm{J/m^2}$ when $A=10^{-11}\,\mathrm{J/m}$ and $K=5\times10^{5}\,\mathrm{J/m^3}$.

\section{Excitations}
\label{appb}

We defined two operators, $\hat{H}_{\phi}$ and $\hat{H}_{\theta}$, in \text{Sec.~\ref{sec:sw}}. In this section, we investigate the properties of the eigenfunctions in detail. The eigenvalue equation of $\hat{H}_{\phi}$ is given by

\begin{equation}
    \hat{H}_\phi \psi^{\phi}_n=[-\frac{d^2}{dx^2}+\frac{1}{\lambda^2}(2\operatorname{sn}^2{\big(\frac{x}{\lambda\kappa},\kappa\big)}-1)]\psi^{\phi}_n= \epsilon^{\phi}_n\psi^{\phi}_n\,.
\end{equation}
After the substitution $s=x/\lambda\kappa$, we reach

\begin{align}
    &\frac{d^2}{dx^2}\psi^{\phi}_n(x)=[\frac{1}{\lambda^2}(2\operatorname{sn}^2\big(\frac{x}{\lambda\kappa},\kappa\big)-1)-\epsilon^{\phi}_n]\psi^{\phi}_n(x)\\
    &\Rightarrow \frac{d^2}{ds^2}\psi^{\phi}_n(s)=[\kappa^2(2\operatorname{sn}^2(s,\kappa)-1)-\frac{\epsilon_n^{\phi}}{q^2}]\psi^{\phi}_n(s)\,,
\end{align}
which corresponds to the Lam\'{e} equation.

\begin{figure}[t]
    \centering
    \includegraphics[width=0.8\linewidth]{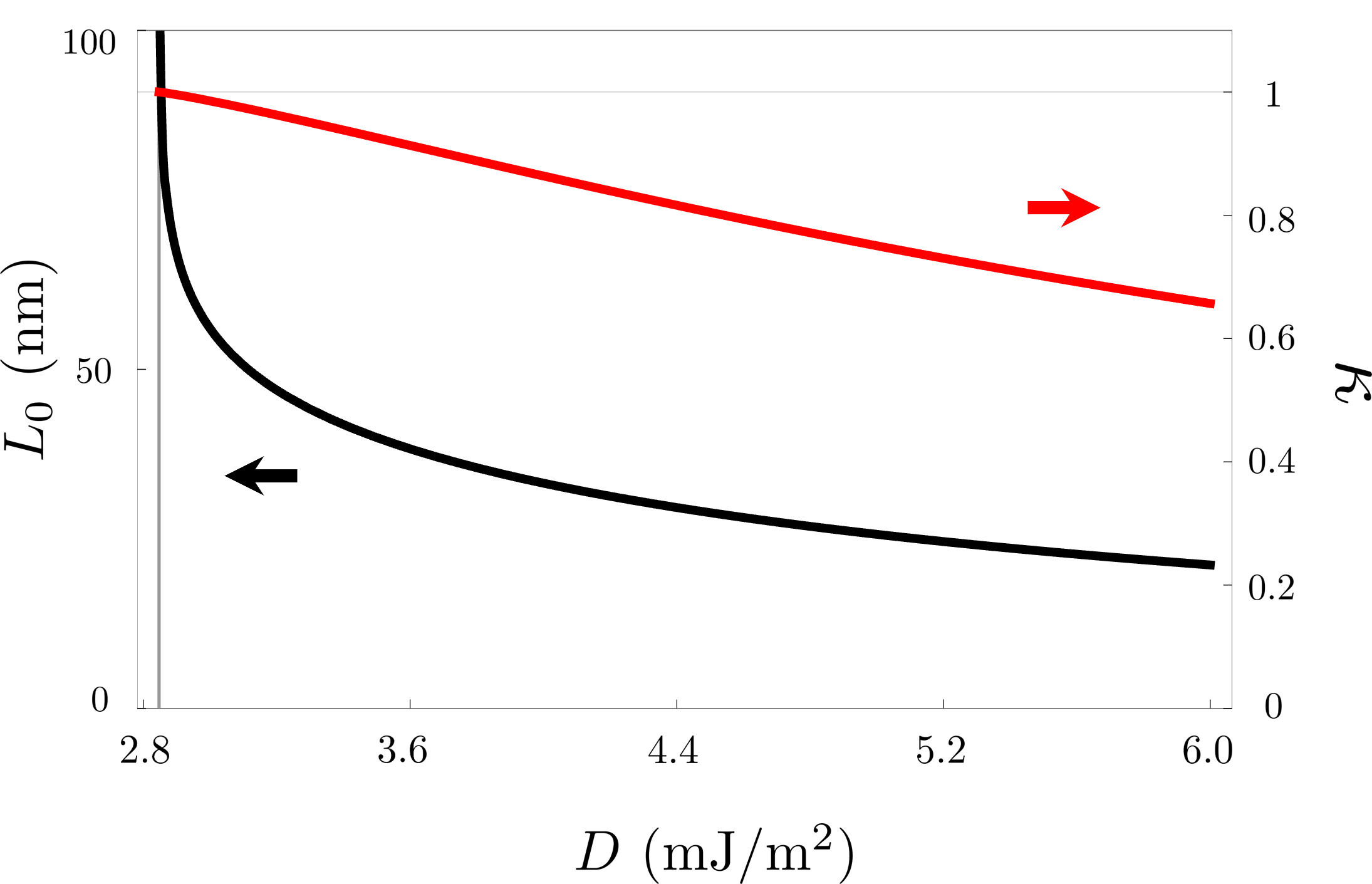} 
    \caption{The black solid line shows the DMI dependence of $L_0$ [Eq.~(\ref{defperiod})] where $A=10^{-11}\,\mathrm{J/m}$ and $K=5\times10^5\,\mathrm{J/m^3}$. The red solid line shows $\kappa$ [Eq.~(\ref{defkappa})], which is the elliptic modulus of Jacobi functions. The vertical line at $D=2.847\,\mathrm{mJ/m^2}$ indicates the critical DMI, $D_c$ [Eq.~(\ref{criticaldmi})].}
    \label{fig:period}
\end{figure}

The solution of the equation is given by

\begin{equation}
    \psi_{\bar{\sigma}}(s)=\frac{H(s-\bar{\sigma})}{\Theta(s)}e^{sZ(\bar{\sigma},\kappa)}\,,
\end{equation}
where $H(s)=\vartheta_1(v, \bar{q})$ is Jacobi eta function with $v=\pi s/2K(\kappa)$ and the elliptic nome $\bar{q}=\exp{[-\pi K'(\kappa)/K(\kappa)]}$, $K(\kappa)$ and $K'(\kappa)=K(\kappa')$ are complete elliptic integrals of the first kind, $\Theta(s)=\vartheta_4(v,\bar{q})$ is Jacobi theta function and $Z(x,\kappa)=\varepsilon(x,\kappa)-\frac{E(\kappa)}{K(\kappa)}x$ is Jacobi zeta function where Jacobi epsilon function is given by $\varepsilon(x,\kappa)=\int_{0}^{x}dt\,\operatorname{dn}^2(t,\kappa)$. Here, $\bar{\sigma}$ was given in Eqs.~(\ref{acoustic}) and (\ref{optical}) for each branch. 

The functions, $H(s)$ and $\Theta(s)$, have the quasi-periodicity as
\begin{gather}
    H[s+2K(\kappa)-\bar{\sigma}]=-H(s-\bar{\sigma})\,,\\
    \Theta[s+2K(\kappa)]=\Theta(s)\,,
\end{gather}
and we can derive the relation,
\begin{equation}
    \psi_{\bar{\sigma}}[s+2K(\kappa)]=-e^{2K(\kappa)Z(\bar{\sigma},\kappa)}\psi_{\bar{\sigma}}(s)\label{bloch1}\,.
\end{equation}

From Bloch's theorem, $Z(\bar{\sigma})$ is purely imaginary in the regimes defined by Eqs.~(\ref{acoustic}) and (\ref{optical}). By introducing a quasi-momentum variable, 
\begin{equation}
    \bar{q}(\bar{\sigma})=\frac{\pi}{2K(\kappa)}+iZ(\bar{\sigma},\kappa)\,,
    \label{quasi}
\end{equation}
we can rewrite Eq.~(\ref{bloch1}) as 
\begin{equation}
    \psi_{\bar{\sigma}}[s+2K(\kappa)]=e^{-2i K(\kappa)\bar{q}(\bar{\sigma})}\psi_{\bar{\sigma}}(s)\,.
\end{equation}

Finally, we can find modes with the parameter, $\bar{\sigma}$, satisfying the periodic boundary condition,
\begin{equation}
    \psi_{\bar{\sigma}}[s+R]=e^{-iR\bar{q}(\bar{\sigma})}\psi_{\bar{\sigma}}[s]=\psi_{\bar{\sigma}}[s]\quad\Rightarrow\quad \bar{q}(\bar{\sigma})=\frac{2\pi}{R}n\,,
\end{equation}
where n is an integer~\cite{kishine2015theory}. Next, we need to choose an appropriate period, $R$. In this paper, the AC magnetic field is uniformly applied over the entire system, not locally. Thus, due to the symmetry, it is reasonable to set $R=L_0$, where $L_0$ is the period of the CSS. 

Based on the condition, we find two modes in the acoustic branch. One of them is the zero mode associated with the spontaneously broken translational symmetry and the other one is the first excited mode, which moves the solitons in a way similar to an optical phonon mode. There exists an energy gap between the two branches. The eigenenergy and the energy gap, $\delta$, at the boundary of the Brillouin zone are given by~\cite{kishine2015theory}

\begin{gather}
    \epsilon_n^{\phi}=q^2\operatorname{dn}^2(\bar{\sigma},\kappa)=
    \begin{cases} 
        q^2 \kappa'^2 \operatorname{sn}^2(\sigma,\kappa')\,, & \text{(acoustic)}\\
        q^2\operatorname{sn}^{-2}(\sigma,\kappa')\,, & \text{(optical)}
    \end{cases}\\
    \delta=A(q^2\kappa^2)=\frac{A}{\lambda^2}=K\,.
    \label{eigenenergy}
\end{gather}
The energy gap only depends on $K$, and the eigenenergy increases with $D$ in both branches because $\kappa$ is inversely proportional to $D$ [See Fig.~\ref{fig:period}].

As mentioned before, there is no exact solution for the eigenvalue equation of $\hat{H}_\theta$, 

\begin{equation}
    \hat{H}_\theta \psi^{\theta}_n=[-\frac{d^2}{dx^2}+\frac{1}{\lambda^2}(2\operatorname{sn}^2{\big(\frac{x}{\lambda\kappa},\kappa\big)}-1)+\Delta(x)]\psi^{\theta}_n= \epsilon^{\theta}_n\psi^{\theta}_n\,,
\end{equation}
because of $\Delta(x)$ [Eq. (\ref{delta})]. This term originates from the DMI and opens a gap at the origin, which makes it more difficult to excite these modes than those from $\hat{H}_\phi$. Thus, the DMI can be interpreted as an effective easy-plane anisotropy, preventing a deviation from the plane $\theta=\pi/2$. The excitation modes can be obtained numerically under the periodic boundary condition and we use the three lowest-eigenenergy modes in the theory.

\begin{figure}[t]
    \centering
    \includegraphics[width=0.8\linewidth]{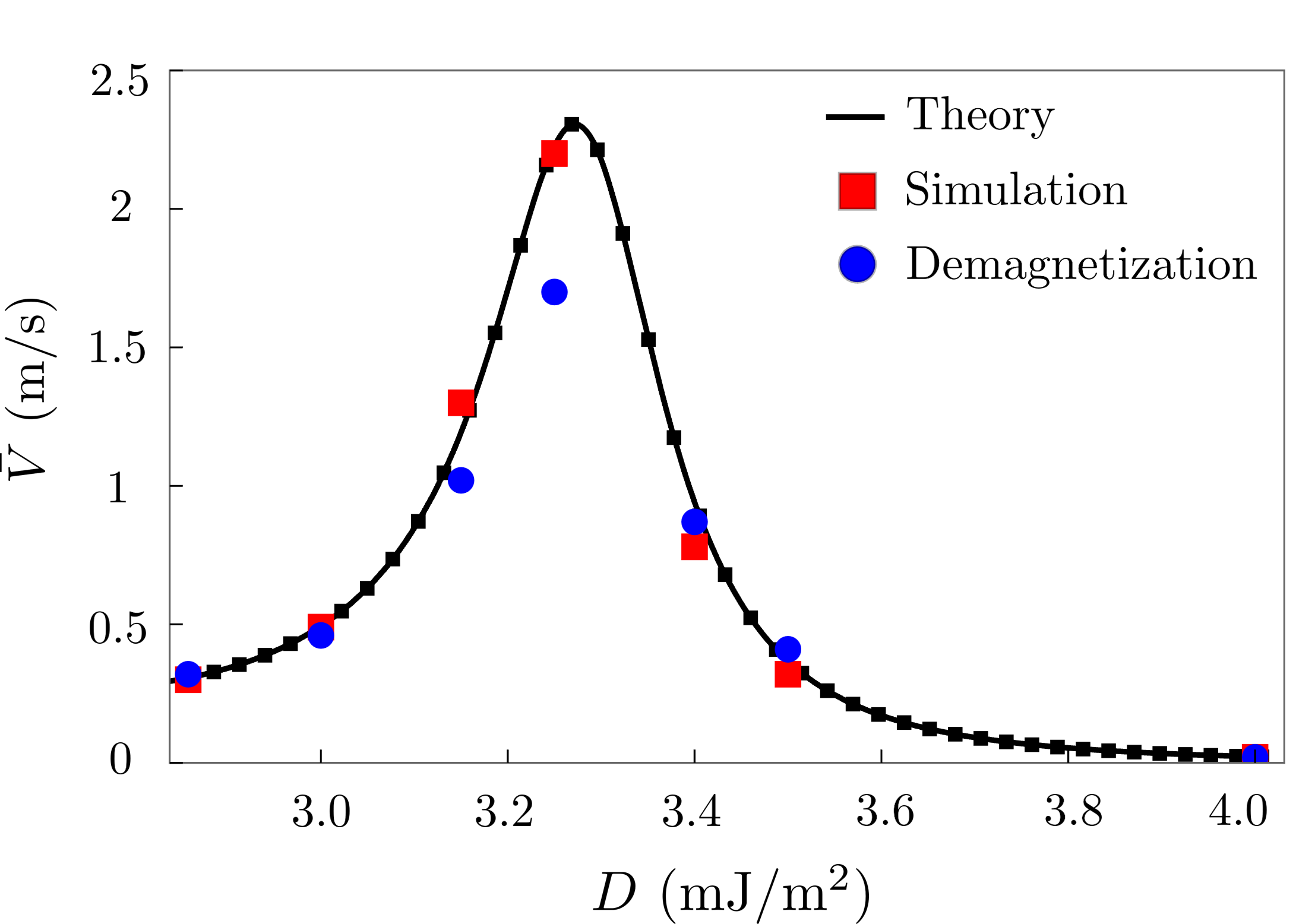} 
    \caption{The plot of DMI dependence of the average velocity, $\bar{V}$ [Eq.~(\ref{avgvel})], at a fixed solid angle, $(\Theta,\Phi)=(\pi/4,0)$, where $f=10\,\mathrm{GHz}$ and $H=30\,\mathrm{mT}$.}
    \label{fig:dmag}
\end{figure}

\section{Dynamics}
\label{appc}

\subsection{The effects of demagnetization}
In practice, ferromagnetic thin films are affected by demagnetization in various respects. One prominent effect is the reduction of easy-axis anisotropy along the film normal. This reduction arises because, in the extremely thin-film limit where dipolar couplings become effectively local, magnetization aligned with the easy-axis induces a demagnetizing field, $\mathbf{H}_d=-(\mathbf{M}\cdot\hat{\mathbf{z}})\hat{\mathbf{z}}$, which opposes both the magnetization and the effective field generated by the anisotropy~\cite{rohart2013skyrmion,blundell2001magnetism}. Here, we incorporate the effect into an effective anisotropy, defined as $K_{\mathrm{eff}}=K-\mu_0 M^2/2$ with the free space permeability, $\mu_0$, and investigate the dynamical effect of demagnetization. We compare the average velocity, $\bar{V}$, in two cases: one with the effective anisotropy accounting for demagnetization (disk symbols), and the other with the same easy-axis anisotropy but without demagnetization (square symbols), as shown in Fig.~\ref{fig:dmag}. The comparison demonstrates that the dynamical effects of demagnetization on $\bar{V}$ result in a marginal decrease in the average velocity.

\vspace{1em}

\subsection{Coefficients}
The coefficients in the Lagrangian and the Rayleigh dissipation function are given below:

\begin{equation}
    G_y=\frac{\mathcal{J}}{\lambda\kappa}\int dx\,\operatorname{dn}u_3\,,\quad\quad
    G_z=-\mathcal{J}\int dx\,u_4u_1\,,    
\end{equation}
\begin{equation}
    G_x=-\mathcal{J}\int dx\,u_5u_2\,,
\end{equation}
\begin{equation}
    C_z=\int dx\,[Au_1'^2-K(\operatorname{cn}^2-\operatorname{sn}^2)u_1^2]\,,
\end{equation}
\begin{equation}
    C_x=\int dx\,[Au_2'^2-K(\operatorname{cn}^2-\operatorname{sn}^2)u_2^2]\,,
\end{equation}
\begin{equation}
    C_y=\int dx\,[A(u_3'^2-\frac{\operatorname{dn}^2u_3^2}{\lambda^2\kappa^2})+K \operatorname{sn}^2u_3^2+\frac{D}{\lambda\kappa}\operatorname{dn}u_3^2]\,,
\end{equation}
\begin{equation}
    C_z'=\int dx\,[A(u_4'^2-\frac{\operatorname{dn}^2u_4^2}{\lambda^2\kappa^2})+K \operatorname{sn}^2u_4^2+\frac{D}{\lambda\kappa}\operatorname{dn}u_4^2]\,,
\end{equation}
\begin{equation}
    C_x'=\int dx\,[A(u_5'^2-\frac{\operatorname{dn}^2u_5^2}{\lambda^2\kappa^2})+K \operatorname{sn}^2u_5^2+\frac{D}{\lambda\kappa}\operatorname{dn}u_5^2]\,,
\end{equation}
\begin{equation}
    Q_z=M\int dx\, \operatorname{cn}u_1\,,\quad
    Q_x=M\int dx\, \operatorname{sn}u_2\,,
\end{equation}
\begin{equation}
    Q_y=M\int dx\, u_3\,,
\end{equation}

and

\begin{equation}
    D_X=\frac{\alpha \mathcal{J}}{2\lambda^2\kappa^2}\int dx\,\operatorname{dn}^2\,,\quad\quad
    D_z=\frac{\alpha \mathcal{J}}{2}\int dx\,u_1^2\,,
\end{equation}
\begin{equation}
    D_x=\frac{\alpha \mathcal{J}}{2}\int dx\,u_2^2\,,\quad\quad
    D_y=\frac{\alpha \mathcal{J}}{2}\int dx\,u_3^2\,,
\end{equation}
\begin{equation}
    {D_z'}=\frac{\alpha \mathcal{J}}{2}\int dx\,u_4^2\,,\quad\quad
    {D_x'}=\frac{\alpha \mathcal{J}}{2}\int dx\,u_5^2\,,
\end{equation}
\begin{equation}
    D_{xz(zx)}=\mp\alpha \mathcal{J}\int dx\,u'_2u_1\,,
\end{equation}
\begin{equation}
    D'_{xz(zx)}=\mp\alpha \mathcal{J}\int dx\,u'_5u_4\,,
\end{equation}
where the integration is performed over one period along the $x$-direction, $u'_i=\partial_x u_i$, and omits $(x/\lambda\kappa,\kappa)$.

\begin{widetext}
\subsection{Solution to EOMs}

The effective solution of EOMs under the assumptions of weak perturbations and the sinusoidal ansatz is given by  

    \begin{equation}
        \tan{\delta_z}\simeq\frac{\omega(4 {C_z'}^2D_z+\omega^2(G_z^2{D_z'}+4D_z{D_z'}^2))}{4 C_z{C_z'}^2-\omega^2(G_z^2C_z'-4C_z{D_z'}^2)}\,,\quad  \tan{\delta_x}\simeq\frac{\omega(4{C_x'}^2D_x+\omega^2(G_x^2{D_x'}+4D_x{D_x'}^2))}{4 C_x{C_x'}^2-\omega^2(G_x^2C_x'-4C_x{D_x'}^2)}\,,
        \label{wide1}
    \end{equation}
    \begin{equation}
        \tan{\delta_y}\simeq\frac{\omega(G_y^2+4D_XD_y)}{4C_yD_X}\,,\quad \tan{\delta_z'}\simeq-\frac{4C_zC_z'-\omega^2(G_z^2+4D_z{D_z'})}{4\omega(C_z'D_z+C_z{D_z'})}\,,\quad \tan{\delta_x'}\simeq-\frac{4C_xC_x'-\omega^2(G_x^2+4D_x{D_x'})}{4\omega(C_x'D_x+C_x{D_x'})}\,,
        \label{wide3}
    \end{equation}
    \begin{equation}
        \tilde{\eta}_z\simeq-\frac{2 Q_z H_z \omega  \csc{\delta_z} \left(4 {C_z'}^2 D_z+{D_z'} \omega ^2 \left(G_z^2+4 D_z
        {D_z'}\right)\right)}{16 C_z^2 {C_z'}^2+8 \omega ^2 \left(-G_z^2C_zC_z'+2 C_z^2 {D_z'}^2+2
        {C_z'}^2 D_z^2\right)+\omega ^4 \left(G_z^2+4 D_z {D_z'}\right)^2}\,,
    \end{equation}
    \begin{equation}
        \tilde{\eta}_x\simeq-\frac{2 Q_x H_x \omega  \csc{\delta_x} \left(4 {C_x'}^2 D_x+{D_x'} \omega ^2 \left(G_x^2+4 D_x{D_x'}\right)\right)}{16 C_x^2 {C_x'}^2+8 \omega ^2 \left(-G_x^2C_xC_x'+2 C_x^2 {D_x'}^2+2{C_x'}^2 D_x^2\right)+\omega ^4 \left(G_x^2+4 D_x {D_x'}\right)^2}\,,
    \end{equation}
    \begin{equation}
        \tilde{\eta}_y\simeq-\frac{2 Q_y D_X H_y\omega\csc{\delta_y}(G_y^2+4D_XD_y)}{16C_y^2D_X^2+\omega^2(G_y^2+4D_XD_y)^2}\,,
    \end{equation}
    \begin{equation}
        \tilde{\pi}_z\simeq\frac{G_z Q_z H_z \omega  \csc{\delta_z'} \left(4 C_z C_z'-\omega^2\left(G_z^2+4 D_z
        {D_z'}\right) \right)}{16 C_z^2 {C_z'}^2+8 \omega ^2 \left(-G_z^2C_zC_z'+2 C_z^2 {D_z'}^2+2
        {C_z'}^2 D_z^2\right)+\omega ^4 \left(G_z^2+4 D_z {D_z'}\right)^2}\,,
    \end{equation}
    \begin{equation}
        \tilde{\pi}_x\simeq\frac{G_x Q_x H_x \omega  \csc{\delta_x'} \left(4 C_x C_x'-\omega^2\left(G_x^2+4 D_x
        {D_x'}\right) \right)}{16 C_x^2 {C_x'}^2+8 \omega ^2 \left(-G_x^2C_xC_x'+2 C_x^2 {D_x'}^2+2
        {C_x'}^2 D_x^2\right)+\omega ^4 \left(G_x^2+4 D_x {D_x'}\right)^2}\,.
    \end{equation}
\end{widetext}

\bibliographystyle{apsrev4-2}
\bibliography{Proj04ref}

@article{kishine_coherent_2012,
	title = {Coherent sliding dynamics and spin motive force driven by crossed magnetic fields in a chiral helimagnet},
	volume = {86},
	copyright = {http://link.aps.org/licenses/aps-default-license},
	issn = {1098-0121, 1550-235X},
	url = {https://link.aps.org/doi/10.1103/PhysRevB.86.214426},
	doi = {10.1103/PhysRevB.86.214426},
	language = {en},
	number = {21},
	urldate = {2024-07-18},
	journal = {Phys. Rev. B},
	author = {Kishine, Jun-ichiro and Bostrem, I. G. and Ovchinnikov, A. S. and Sinitsyn, Vl. E.},
	month = dec,
	year = {2012},
	pages = {214426},
  URL={https://journals.aps.org/prb/abstract/10.1103/PhysRevB.86.214426}
}

@article{moon2017domain,
  title={Domain wall motion driven by an oscillating magnetic field},
  author={Moon, Kyoung-Woong and Kim, Duck-Ho and Kim, Changsoo and Kim, Dae-Yun and Choe, Sug-Bong and Hwang, Chanyong},
  journal={	J. Phys. D: Appl. Phys.},
  volume={50},
  number={12},
  pages={125003},
  year={2017},
  publisher={IOP Publishing},
  URL={https://iopscience.iop.org/article/10.1088/1361-6463/aa5d35/meta}
}

@article{beach2005dynamics,
  title={Dynamics of field-driven domain-wall propagation in ferromagnetic nanowires},
  author={Beach, Geoffrey SD and Nistor, Corneliu and Knutson, Carl and Tsoi, Maxim and Erskine, James L},
  journal={Nat. Mater.},
  volume={4},
  number={10},
  pages={741--744},
  year={2005},
  publisher={Nature Publishing Group UK London},
  URL={https://www.nature.com/articles/nmat1477}
}

@article{kim2020magnetic,
  title={Magnetic soliton rectifier via phase synchronization},
  author={Kim, Duck-Ho and Kim, Dong-Hyun and Kim, Dae-Yun and Choe, Sug-Bong and Ono, Teruo and Lee, Kyung-Jin and Kim, Se Kwon},
  journal={Phys. Rev. B},
  volume={102},
  number={18},
  pages={184430},
  year={2020},
  publisher={APS},
  URL={https://journals.aps.org/prb/abstract/10.1103/PhysRevB.102.184430}
}

@article{rohart2013skyrmion,
  title={Skyrmion confinement in ultrathin film nanostructures in the presence of Dzyaloshinskii-Moriya interaction},
  author={Rohart, S and Thiaville, A},
  journal={Phys. Rev. B},
  volume={88},
  number={18},
  pages={184422},
  year={2013},
  publisher={APS},
  URL={https://journals.aps.org/prb/abstract/10.1103/PhysRevB.88.184422}
}

@article{kim2023mechanics,
  title={Mechanics of a ferromagnetic domain wall},
  author={Kim, Se Kwon and Tchernyshyov, Oleg},
  journal={	J. Phys. Condens. Matter},
  volume={35},
  number={13},
  pages={134002},
  year={2023},
  publisher={IOP Publishing},
  URL={https://iopscience.iop.org/article/10.1088/1361-648X/acb5d8/meta}
}

@article{vansteenkiste2014design,
  title={The design and verification of MuMax3},
  author={Vansteenkiste, Arne and Leliaert, Jonathan and Dvornik, Mykola and Helsen, Mathias and Garcia-Sanchez, Felipe and Van Waeyenberge, Bartel},
  journal={AIP Adv.},
  volume={4},
  number={10},
  year={2014},
  publisher={AIP Publishing},
  URL={https://pubs.aip.org/aip/adv/article/4/10/107133/584191}
}

@article{kishine2015theory,
  title={Theory of monoaxial chiral helimagnet},
  author={Kishine, Jun-ichiro and Ovchinnikov, AS},
  journal={Solid State Phys.},
  volume={66},
  pages={1--130},
  year={2015},
  publisher={Elsevier},
  URL={https://www.sciencedirect.com/science/article/abs/pii/S0081194715000041}
}

@article{moriya1960anisotropic,
  title={Anisotropic superexchange interaction and weak ferromagnetism},
  author={Moriya, T{\^o}ru},
  journal={Phys. Rev.},
  volume={120},
  number={1},
  pages={91},
  year={1960},
  publisher={APS},
  URL={https://journals.aps.org/pr/abstract/10.1103/PhysRev.120.91}
}

@article{clarke2008dynamics,
  title={Dynamics of a vortex domain wall in a magnetic nanostrip: Application of the collective-coordinate approach},
  author={Clarke, DJ and Tretiakov, OA and Chern, G-W and Bazaliy, Ya B and Tchernyshyov, O},
  journal={Phys. Rev. B},
  volume={78},
  number={13},
  pages={134412},
  year={2008},
  publisher={APS},
  URL={https://journals.aps.org/prb/abstract/10.1103/PhysRevB.78.134412}
}

@article{schryer1974motion,
  title={The motion of 180 domain walls in uniform dc magnetic fields},
  author={Schryer, Norman L and Walker, Laurence R},
  journal={J. Appl. Phys.},
  volume={45},
  number={12},
  pages={5406--5421},
  year={1974},
  publisher={American Institute of Physics},
  URL={https://pubs.aip.org/aip/jap/article/45/12/5406/168816/The-motion-of-180-domain-walls-in-uniform-dc}
}

@article{barone1971theory,
  title={Theory and applications of the sine-Gordon equation},
  author={Barone, A and Esposito, F and Magee, CJ and Scott, AC},
  journal={Riv. Nuovo Cimento},
  volume={1},
  number={2},
  pages={227--267},
  year={1971},
  publisher={Springer},
  URL={https://link.springer.com/article/10.1007/BF02820622}
}

@article{ghosh2017annihilation,
  title={Annihilation of domain walls in a ferromagnetic wire},
  author={Ghosh, Anirban and Huang, Kevin S and Tchernyshyov, Oleg},
  journal={Phys. Rev. B},
  volume={95},
  number={18},
  pages={180408},
  year={2017},
  publisher={APS},
  URL={https://journals.aps.org/prb/abstract/10.1103/PhysRevB.95.180408}
}

@article{thiaville2012dynamics,
  title={Dynamics of Dzyaloshinskii domain walls in ultrathin magnetic films},
  author={Thiaville, Andr{\'e} and Rohart, Stanislas and Ju{\'e}, {\'E}milie and Cros, Vincent and Fert, Albert},
  journal={EPL},
  volume={100},
  number={5},
  pages={57002},
  year={2012},
  publisher={IOP Publishing},
  URL={https://iopscience.iop.org/article/10.1209/0295-5075/100/57002/meta}
}

@article{slonczewski1972dynamics,
    author = {Slonczewski, J. C.},
    title = {DYNAMICS OF MAGNETIC DOMAIN WALLS},
    journal = {AIP Conf. Proc.},
    volume = {5},
    number = {1},
    pages = {170-174},
    year = {1972},
    month = {03},
    doi = {10.1063/1.3699416},
    url = {https://doi.org/10.1063/1.3699416}
}

@article{dzyaloshinsky1958thermodynamic,
  title={A thermodynamic theory of “weak” ferromagnetism of antiferromagnetics},
  author={Dzyaloshinsky, Igor},
  journal={J. Phys. Chem. Solids},
  volume={4},
  number={4},
  pages={241--255},
  year={1958},
  publisher={Elsevier},
  URL={https://www.sciencedirect.com/science/article/pii/0022369758900763}
}

@article{anderson1959new,
  title={New approach to the theory of superexchange interactions},
  author={Anderson, Philip W},
  journal={	Phys. Rev.},
  volume={115},
  number={1},
  pages={2},
  year={1959},
  publisher={APS},
  URL={https://journals.aps.org/pr/abstract/10.1103/PhysRev.115.2}
}

@article{PhysRevLett.103.157201,
  title = {Real-Space Observation of a Right-Rotating Inhomogeneous Cycloidal Spin Spiral by Spin-Polarized Scanning Tunneling Microscopy in a Triple Axes Vector Magnet},
  author = {Meckler, S. and Mikuszeit, N. and Pre\ss{}ler, A. and Vedmedenko, E. Y. and Pietzsch, O. and Wiesendanger, R.},
  journal = {Phys. Rev. Lett.},
  volume = {103},
  issue = {15},
  pages = {157201},
  numpages = {4},
  year = {2009},
  month = {Oct},
  publisher = {American Physical Society},
  doi = {10.1103/PhysRevLett.103.157201},
  url = {https://link.aps.org/doi/10.1103/PhysRevLett.103.157201}
}

@article{PhysRevLett.101.027201,
  title = {Atomic-Scale Spin Spiral with a Unique Rotational Sense: Mn Monolayer on W(001)},
  author = {Ferriani, P. and von Bergmann, K. and Vedmedenko, E. Y. and Heinze, S. and Bode, M. and Heide, M. and Bihlmayer, G. and Bl\"ugel, S. and Wiesendanger, R.},
  journal = {Phys. Rev. Lett.},
  volume = {101},
  issue = {2},
  pages = {027201},
  numpages = {4},
  year = {2008},
  month = {Jul},
  publisher = {American Physical Society},
  doi = {10.1103/PhysRevLett.101.027201},
  url = {https://link.aps.org/doi/10.1103/PhysRevLett.101.027201}
}

@article{bode2007chiral,
  title={Chiral magnetic order at surfaces driven by inversion asymmetry},
  author={Bode, Matthias and Heide, M and Von Bergmann, K and Ferriani, P and Heinze, Stefan and Bihlmayer, G and Kubetzka, A and Pietzsch, O and Bl{\"u}gel, Stefan and Wiesendanger, R},
  journal={Nature},
  volume={447},
  number={7141},
  pages={190--193},
  year={2007},
  publisher={Nature Publishing Group UK London},
  URL={https://www.nature.com/articles/nature05802}
}

@article{kenzelmann2005magnetic,
  title={Magnetic inversion symmetry breaking and ferroelectricity in TbMnO 3},
  author={Kenzelmann, Michel and Harris, A Brooks and Jonas, S and Broholm, C and Schefer, J{\"u}rg and Kim, SB and Zhang, CL and Cheong, <? format?> S-W and Vajk, Owen P and Lynn, Jeffrey W},
  journal={Phys. Rev. Lett.},
  volume={95},
  number={8},
  pages={087206},
  year={2005},
  publisher={APS},
  URL={https://journals.aps.org/prl/abstract/10.1103/PhysRevLett.95.087206}
}

@article{PhysRevLett.106.047203,
  title = {Coupled Magnetic Cycloids in Multiferroic ${\mathrm{TbMnO}}_{3}$ and ${\mathrm{Eu}}_{3/4}{\mathrm{Y}}_{1/4}{\mathrm{MnO}}_{3}$},
  author = {Jang, Hoyoung and Lee, J.-S. and Ko, K.-T. and Noh, W.-S. and Koo, T. Y. and Kim, J.-Y. and Lee, K.-B. and Park, J.-H. and Zhang, C. L. and Kim, Sung Baek and Cheong, S.-W.},
  journal = {Phys. Rev. Lett.},
  volume = {106},
  issue = {4},
  pages = {047203},
  numpages = {4},
  year = {2011},
  month = {Jan},
  publisher = {American Physical Society},
  doi = {10.1103/PhysRevLett.106.047203},
  url = {https://link.aps.org/doi/10.1103/PhysRevLett.106.047203}
}

@article{Park_2014,
doi = {10.1088/0953-8984/26/43/433202},
url = {https://dx.doi.org/10.1088/0953-8984/26/43/433202},
year = {2014},
month = {oct},
publisher = {IOP Publishing},
volume = {26},
number = {43},
pages = {433202},
author = {Park, Je-Geun and Le, Manh Duc and Jeong, Jaehong and Lee, Sanghyun},
title = {Structure and spin dynamics of multiferroic BiFeO3},
journal = {J. Phys. Condens. Matter}
}

@article{PhysRevB.84.144404,
  title = {Low-temperature evolution of the modulated magnetic structure in the ferroelectric antiferromagnet BiFeO${}_{3}$},
  author = {Sosnowska, I. and Przenios\l{}o, R.},
  journal = {Phys. Rev. B},
  volume = {84},
  issue = {14},
  pages = {144404},
  numpages = {5},
  year = {2011},
  month = {Oct},
  publisher = {American Physical Society},
  doi = {10.1103/PhysRevB.84.144404},
  url = {https://link.aps.org/doi/10.1103/PhysRevB.84.144404}
}

@article{thiaville2002domain,
  title={Domain wall dynamics in nanowires},
  author={Thiaville, A and Garc{\i}a, JM and Miltat, J},
  journal={J. Magn. Magn. Mater.},
  volume={242},
  pages={1061--1063},
  year={2002},
  publisher={Elsevier},
  URL={https://www.sciencedirect.com/science/article/pii/S0304885301013531}
}

@article{yoshimori,
author = {Yoshimori ,Akio},
title = {A New Type of Antiferromagnetic Structure in the Rutile Type Crystal},
journal = {J. Phys. Soc. Jpn.},
volume = {14},
number = {6},
pages = {807-821},
year = {1959},
doi = {10.1143/JPSJ.14.807},

URL = { 
        https://doi.org/10.1143/JPSJ.14.807
},
}

@article{nagamiya1962magnetization,
  title={Magnetization process of a screw spin system},
  author={Nagamiya, Takeo and Nagata, Kazukiyo and Kitano, Yoshiharu},
  journal={	Prog. Theor. Exp. Phys.},
  volume={27},
  number={6},
  pages={1253--1271},
  year={1962},
  publisher={Oxford University Press},
  URL = { 
        https://academic.oup.com/ptp/article/27/6/1253/1907149?login=true
  },
}

@article{LIU2018495,
title = {Vortical structures for nanomagnetic memory induced by dipole-dipole interaction in monolayer disks},
journal = {Superlattice. Microst.},
volume = {117},
pages = {495-502},
year = {2018},
issn = {0749-6036},
doi = {https://doi.org/10.1016/j.spmi.2018.03.058},
url = {https://www.sciencedirect.com/science/article/pii/S0749603618301526},
author = {Zhaosen Liu and Orion Ciftja and Xichao Zhang and Yan Zhou and Hou Ian}
}

@article{RevModPhys.91.015004,
  title = {Magnetic small-angle neutron scattering},
  author = {M\"uhlbauer, Sebastian and Honecker, Dirk and P\'erigo, \'Elio A. and Bergner, Frank and Disch, Sabrina and Heinemann, Andr\'e and Erokhin, Sergey and Berkov, Dmitry and Leighton, Chris and Eskildsen, Morten Ring and Michels, Andreas},
  journal = {Rev. Mod. Phys.},
  volume = {91},
  issue = {1},
  pages = {015004},
  numpages = {75},
  year = {2019},
  month = {Mar},
  publisher = {American Physical Society},
  doi = {10.1103/RevModPhys.91.015004},
  url = {https://link.aps.org/doi/10.1103/RevModPhys.91.015004}
}

@article{PhysRevLett.31.459,
  title = {Effect of Antisymmetric Interactions on Critical Phenomena: A System with Helical Ground State},
  author = {Liu, Luke L.},
  journal = {Phys. Rev. Lett.},
  volume = {31},
  issue = {7},
  pages = {459--462},
  numpages = {0},
  year = {1973},
  month = {Aug},
  publisher = {American Physical Society},
  doi = {10.1103/PhysRevLett.31.459},
  url = {https://link.aps.org/doi/10.1103/PhysRevLett.31.459}
}

@book{chaikin1995principles,
  title={Principles of condensed matter physics},
  author={Chaikin, Paul M and Lubensky, Tom C and Witten, Thomas A},
  volume={10},
  year={1995},
  publisher={Cambridge university press Cambridge}
}

@article{poschl1933bemerkungen,
  title={Bemerkungen zur Quantenmechanik des anharmonischen Oszillators},
  author={P{\"o}schl, Gertrud and Teller, Edward},
  journal={Z. Phys.},
  volume={83},
  number={3},
  pages={143--151},
  year={1933},
  publisher={Springer},
  URL={https://link.springer.com/article/10.1007/BF01331132}
}

@article{PhysRevB.78.140403,
  title = {Dzyaloshinskii-Moriya interaction accounting for the orientation of magnetic domains in ultrathin films: Fe/W(110)},
  author = {Heide, M. and Bihlmayer, G. and Bl\"ugel, S.},
  journal = {Phys. Rev. B},
  volume = {78},
  issue = {14},
  pages = {140403},
  numpages = {4},
  year = {2008},
  month = {Oct},
  publisher = {American Physical Society},
  doi = {10.1103/PhysRevB.78.140403},
  url = {https://link.aps.org/doi/10.1103/PhysRevB.78.140403}
}

@article{PhysRevB.104.224405,
  title = {Spin excitation spectra in helimagnetic states: Proper-screw, cycloid, vortex-crystal, and hedgehog lattices},
  author = {Kato, Yasuyuki and Hayami, Satoru and Motome, Yukitoshi},
  journal = {Phys. Rev. B},
  volume = {104},
  issue = {22},
  pages = {224405},
  numpages = {21},
  year = {2021},
  month = {Dec},
  publisher = {American Physical Society},
  doi = {10.1103/PhysRevB.104.224405},
  url = {https://link.aps.org/doi/10.1103/PhysRevB.104.224405}
}

@article{sharma2024nanoscale,
  title={Nanoscale spin rectifiers for harvesting ambient radiofrequency energy},
  author={Sharma, Raghav and Ngo, Tung and Raimondo, Eleonora and Giordano, Anna and Igarashi, Junta and Jinnai, Butsurin and Zhao, Shishun and Lei, Jiayu and Guo, Yong-Xin and Finocchio, Giovanni and others},
  journal={Nat. Electron.},
  volume={7},
  number={8},
  pages={653--661},
  year={2024},
  publisher={Nature Publishing Group UK London},
  URL={https://www.nature.com/articles/s41928-024-01212-1}
}

@article{PhysRevB.77.134407,
  title = {Electron transport driven by nonequilibrium magnetic textures},
  author = {Tserkovnyak, Yaroslav and Mecklenburg, Matthew},
  journal = {Phys. Rev. B},
  volume = {77},
  issue = {13},
  pages = {134407},
  numpages = {4},
  year = {2008},
  month = {Apr},
  publisher = {American Physical Society},
  doi = {10.1103/PhysRevB.77.134407},
  url = {https://link.aps.org/doi/10.1103/PhysRevB.77.134407}
}

@article{PhysRevLett.134.056702,
  title = {Current-Induced Sliding Motion in a Helimagnet ${\mathrm{MnAu}}_{2}$},
  author = {Kimoto, Yuta and Masuda, Hidetoshi and Seki, Takeshi and Nii, Yoichi and Ohe, Jun-ichiro and Nambu, Yusuke and Onose, Yoshinori},
  journal = {Phys. Rev. Lett.},
  volume = {134},
  issue = {5},
  pages = {056702},
  numpages = {6},
  year = {2025},
  month = {Feb},
  publisher = {American Physical Society},
  doi = {10.1103/PhysRevLett.134.056702},
  url = {https://link.aps.org/doi/10.1103/PhysRevLett.134.056702}
}

@article{PhysRevB.82.054410,
  title = {Topological electromotive force from domain-wall dynamics in a ferromagnet},
  author = {Yang, Shengyuan A. and Beach, Geoffrey S. D. and Knutson, Carl and Xiao, Di and Zhang, Zhenyu and Tsoi, Maxim and Niu, Qian and MacDonald, A. H. and Erskine, James L.},
  journal = {Phys. Rev. B},
  volume = {82},
  issue = {5},
  pages = {054410},
  numpages = {12},
  year = {2010},
  month = {Aug},
  publisher = {American Physical Society},
  doi = {10.1103/PhysRevB.82.054410},
  url = {https://link.aps.org/doi/10.1103/PhysRevB.82.054410}
}

@article{PhysRevB.95.184419,
  title = {Effect of spin transfer torque on domain wall motion regimes in [Co/Ni] superlattice wires},
  author = {Le Gall, S. and Vernier, N. and Montaigne, F. and Thiaville, A. and Sampaio, J. and Ravelosona, D. and Mangin, S. and Andrieu, S. and Hauet, T.},
  journal = {Phys. Rev. B},
  volume = {95},
  issue = {18},
  pages = {184419},
  numpages = {7},
  year = {2017},
  month = {May},
  publisher = {American Physical Society},
  doi = {10.1103/PhysRevB.95.184419},
  url = {https://link.aps.org/doi/10.1103/PhysRevB.95.184419}
}

@article{pollard2012direct,
  title={Direct dynamic imaging of non-adiabatic spin torque effects},
  author={Pollard, SD and Huang, L and Buchanan, KS and Arena, DA and Zhu, Y},
  journal={Nat. Commun.},
  volume={3},
  number={1},
  pages={1028},
  year={2012},
  publisher={Nature Publishing Group UK London},
  URL={https://www.nature.com/articles/ncomms2025}
}

@book{blundell2001magnetism,
  title={Magnetism in condensed matter},
  author={Blundell, Stephen},
  year={2001},
  publisher={OUP Oxford}
}

@article{PhysRevLett.93.127204,
  title = {Roles of Nonequilibrium Conduction Electrons on the Magnetization Dynamics of Ferromagnets},
  author = {Zhang, S. and Li, Z.},
  journal = {Phys. Rev. Lett.},
  volume = {93},
  issue = {12},
  pages = {127204},
  numpages = {4},
  year = {2004},
  month = {Sep},
  publisher = {American Physical Society},
  doi = {10.1103/PhysRevLett.93.127204},
  url = {https://link.aps.org/doi/10.1103/PhysRevLett.93.127204}
}

@book{landau2013course,
  title={Course of theoretical physics},
  author={Landau, Lev Davidovich and Lifshitz, Evgenii Mikhailovich},
  year={2013},
  publisher={Elsevier}
}

@article{PhysRevLett.108.217202,
  title = {Prediction of Giant Spin Motive Force due to Rashba Spin-Orbit Coupling},
  author = {Kim, Kyoung-Whan and Moon, Jung-Hwan and Lee, Kyung-Jin and Lee, Hyun-Woo},
  journal = {Phys. Rev. Lett.},
  volume = {108},
  issue = {21},
  pages = {217202},
  numpages = {5},
  year = {2012},
  month = {May},
  publisher = {American Physical Society},
  doi = {10.1103/PhysRevLett.108.217202},
  url = {https://link.aps.org/doi/10.1103/PhysRevLett.108.217202}
}

@article{PhysRevB.87.054403,
  title = {Spin motive force induced by Rashba interaction in the strong $sd$ coupling regime},
  author = {Tatara, Gen and Nakabayashi, Noriyuki and Lee, Kyung-Jin},
  journal = {Phys. Rev. B},
  volume = {87},
  issue = {5},
  pages = {054403},
  numpages = {9},
  year = {2013},
  month = {Feb},
  publisher = {American Physical Society},
  doi = {10.1103/PhysRevB.87.054403},
  url = {https://link.aps.org/doi/10.1103/PhysRevB.87.054403}
}

@article{PhysRevB.91.180405,
  title = {Interfacial Dzyaloshinskii-Moriya interaction in perpendicularly magnetized ${\text{Pt/Co/AlO}}_{x}$ ultrathin films measured by Brillouin light spectroscopy},
  author = {Belmeguenai, Mohamed and Adam, Jean-Paul and Roussign\'e, Yves and Eimer, Sylvain and Devolder, Thibaut and Kim, Joo-Von and Cherif, Salim Mourad and Stashkevich, Andrey and Thiaville, Andr\'e},
  journal = {Phys. Rev. B},
  volume = {91},
  issue = {18},
  pages = {180405},
  numpages = {4},
  year = {2015},
  month = {May},
  publisher = {American Physical Society},
  doi = {10.1103/PhysRevB.91.180405},
  url = {https://link.aps.org/doi/10.1103/PhysRevB.91.180405}
}

@article{RevModPhys.89.025006,
  title = {Interface-induced phenomena in magnetism},
  author = {Hellman, Frances and Hoffmann, Axel and Tserkovnyak, Yaroslav and Beach, Geoffrey S. D. and Fullerton, Eric E. and Leighton, Chris and MacDonald, Allan H. and Ralph, Daniel C. and Arena, Dario A. and D\"urr, Hermann A. and Fischer, Peter and Grollier, Julie and Heremans, Joseph P. and Jungwirth, Tomas and Kimel, Alexey V. and Koopmans, Bert and Krivorotov, Ilya N. and May, Steven J. and Petford-Long, Amanda K. and Rondinelli, James M. and Samarth, Nitin and Schuller, Ivan K. and Slavin, Andrei N. and Stiles, Mark D. and Tchernyshyov, Oleg and Thiaville, Andr\'e and Zink, Barry L.},
  journal = {Rev. Mod. Phys.},
  volume = {89},
  issue = {2},
  pages = {025006},
  numpages = {79},
  year = {2017},
  month = {Jun},
  publisher = {American Physical Society},
  doi = {10.1103/RevModPhys.89.025006},
  url = {https://link.aps.org/doi/10.1103/RevModPhys.89.025006}
}

@article{nasseri2018collective,
  title={Collective coordinate descriptions of magnetic domain wall motion in perpendicularly magnetized nanostructures under the application of in-plane fields},
  author={Nasseri, S Ali and Martinez, Eduardo and Durin, Gianfranco},
  journal={J. Magn. Magn. Mater.},
  volume={468},
  pages={25--43},
  year={2018},
  publisher={Elsevier},
  URL={https://www.sciencedirect.com/science/article/pii/S0304885318309259}
}

@article{RevModPhys.95.015003,
  title = {Measuring interfacial Dzyaloshinskii-Moriya interaction in ultrathin magnetic films},
  author = {Kuepferling, M. and Casiraghi, A. and Soares, G. and Durin, G. and Garcia-Sanchez, F. and Chen, L. and Back, C. H. and Marrows, C. H. and Tacchi, S. and Carlotti, G.},
  journal = {Rev. Mod. Phys.},
  volume = {95},
  issue = {1},
  pages = {015003},
  numpages = {56},
  year = {2023},
  month = {Mar},
  publisher = {American Physical Society},
  doi = {10.1103/RevModPhys.95.015003},
  url = {https://link.aps.org/doi/10.1103/RevModPhys.95.015003}
}

@article{ryu2012current,
  title={Current induced tilting of domain walls in high velocity motion along perpendicularly magnetized micron-sized Co/Ni/Co racetracks},
  author={Ryu, Kwang-Su and Thomas, Luc and Yang, See-Hun and Parkin, Stuart SP},
  journal={Appl. Phys. Express},
  volume={5},
  number={9},
  pages={093006},
  year={2012},
  publisher={IOP Publishing},
  URL={https://iopscience.iop.org/article/10.1143/APEX.5.093006/meta}
}

@article{PhysRevLett.111.217203,
  title = {Domain Wall Tilting in the Presence of the Dzyaloshinskii-Moriya Interaction in Out-of-Plane Magnetized Magnetic Nanotracks},
  author = {Boulle, O. and Rohart, S. and Buda-Prejbeanu, L. D. and Ju\'e, E. and Miron, I. M. and Pizzini, S. and Vogel, J. and Gaudin, G. and Thiaville, A.},
  journal = {Phys. Rev. Lett.},
  volume = {111},
  issue = {21},
  pages = {217203},
  numpages = {5},
  year = {2013},
  month = {Nov},
  publisher = {American Physical Society},
  doi = {10.1103/PhysRevLett.111.217203},
  url = {https://link.aps.org/doi/10.1103/PhysRevLett.111.217203}
}

@misc{myfootnote,
  note = {The evaluation of SMF from spin-transfer torque is also conducted. However, the magnitude of DC voltage is on the order of picovolts for both adiabatic and non-adiabatic contributions, which is negligible.}
}

@misc{note2,
  note = {As $K$ decreases with constant $D$, the gap between acoustic and optical branches narrows, and the two resonance frequencies merge at a specific frequency. For the case of $A=10^{-11}\,\mathrm{J/m}$, $K=5\times10^4\,\mathrm{J/m^3}$, and $D=3.4\, \mathrm{mJ/m^2}$, the resonance occurs at $f=23 \,\mathrm{GHz}$, where the average velocity reaches $12\,\mathrm{m/s}$ under an AC magnetic field of solid angle $(\Theta,\Phi)=(\pi/4,0)$ and amplitude $H=30\,\mathrm{mT}$. Furthermore, we confirmed that the CSS attains approximately $1\,\mathrm{m/s}$ under a field of $10\,\mathrm{Oe}$ when $D=0.5\, \mathrm{mJ/m^2}$.}
}

@article{KOSEVICH1990117,
title = {Magnetic Solitons},
journal = {	Phys. Rep.},
volume = {194},
number = {3},
pages = {117-238},
year = {1990},
issn = {0370-1573},
doi = {https://doi.org/10.1016/0370-1573(90)90130-T},
url = {https://www.sciencedirect.com/science/article/pii/037015739090130T},
author = {A.M. Kosevich and B.A. Ivanov and A.S. Kovalev}
}

@article{Braun01022012,
author = {Hans-Benjamin Braun},
title = {Topological effects in nanomagnetism: from superparamagnetism to chiral quantum solitons},
journal = {Adv. Phys.},
volume = {61},
number = {1},
pages = {1--116},
year = {2012},
publisher = {Taylor \& Francis},
doi = {10.1080/00018732.2012.663070},
URL = { 
        https://doi.org/10.1080/00018732.2012.663070
}
}

@article{Dzyaloshinskii1964rus,
  author  = {I. E. Dzyaloshinskii},
  title   = {Theory of helicoidal structures in antiferromagnets},
  journal = {Zh. Eksp. Teor. Fiz.},
  volume  = {47},
  pages   = {992},
  year    = {1964}
}

@article{Yurii,
doi = {10.1070/PU1984v027n11ABEH004120},
url = {https://dx.doi.org/10.1070/PU1984v027n11ABEH004120},
year = {1984},
month = {nov},
publisher = {},
volume = {27},
number = {11},
pages = {845},
author = {Yurii A Izyumov},
title = {Modulated, or long-periodic, magnetic structures of crystals},
journal = {Phys.-Uspekhi}
}

@article{PhysRevLett.57.1488,
  title = {Geometrical Interpretation of Momentum and Crystal Momentum of Classical and Quantum Ferromagnetic Heisenberg Chains},
  author = {Haldane, F. D. M.},
  journal = {Phys. Rev. Lett.},
  volume = {57},
  issue = {12},
  pages = {1488--1491},
  numpages = {0},
  year = {1986},
  month = {Sep},
  publisher = {American Physical Society},
  doi = {10.1103/PhysRevLett.57.1488},
  url = {https://link.aps.org/doi/10.1103/PhysRevLett.57.1488}
}

@article{duine2018synthetic,
  title={Synthetic antiferromagnetic spintronics},
  author={Duine, RA and Lee, Kyung-Jin and Parkin, Stuart SP and Stiles, Mark D},
  journal={	Nat. Phys.},
  volume={14},
  number={3},
  pages={217--219},
  year={2018},
  publisher={Nature Publishing Group UK London},
  url={https://www.nature.com/articles/s41567-018-0050-y}
}

@article{yang2015domain,
  title={Domain-wall velocities of up to 750 m s- 1 driven by exchange-coupling torque in synthetic antiferromagnets},
  author={Yang, See-Hun and Ryu, Kwang-Su and Parkin, Stuart},
  journal={Nat. Nanotechnol.},
  volume={10},
  number={3},
  pages={221--226},
  year={2015},
  publisher={Nature Publishing Group UK London},
  url={https://www.nature.com/articles/nnano.2014.324}
}

@article{PhysRevB.79.024423,
  title = {Influence of thermal annealing on the perpendicular magnetic anisotropy of Pt/Co/AlOx trilayers},
  author = {Rodmacq, B. and Manchon, A. and Ducruet, C. and Auffret, S. and Dieny, B.},
  journal = {Phys. Rev. B},
  volume = {79},
  issue = {2},
  pages = {024423},
  numpages = {8},
  year = {2009},
  month = {Jan},
  publisher = {American Physical Society},
  doi = {10.1103/PhysRevB.79.024423},
  url = {https://link.aps.org/doi/10.1103/PhysRevB.79.024423}
}

@article{PhysRevApplied.13.044029,
  title = {Effects of Oxidation of Top and Bottom Interfaces on the Electric, Magnetic, and Spin-Orbit Torque Properties of $\mathrm{Pt}$/$\mathrm{Co}$/${\mathrm{Al}\mathrm{O}}_{x}$ Trilayers},
  author = {Feng, Junxiao and Grimaldi, Eva and Avci, Can Onur and Baumgartner, Manuel and Cossu, Giovanni and Rossi, Antonella and Gambardella, Pietro},
  journal = {Phys. Rev. Appl.},
  volume = {13},
  issue = {4},
  pages = {044029},
  numpages = {17},
  year = {2020},
  month = {Apr},
  publisher = {American Physical Society},
  doi = {10.1103/PhysRevApplied.13.044029},
  url = {https://link.aps.org/doi/10.1103/PhysRevApplied.13.044029}
}
\end{document}